\documentclass[twocolumn,aps,amsmath,amssymb,superscriptaddress,prx]{revtex4-1}   
\usepackage[latin2]{inputenc}
\usepackage{amsmath,amssymb}
\usepackage{amsfonts}
\usepackage{bm}
\usepackage{dcolumn}
\usepackage{setspace}
\usepackage{graphicx}
\usepackage{color}
\usepackage{multirow}
\usepackage{verbatim}
\usepackage{epstopdf}
\def\bar{\begin{array}}
\def\ear{\end{array}}

\def\u{\uparrow}
\def\d{\downarrow}

\def\s{\sigma}
\def\f{\frac}

\def\nn{\nonumber}
\def\R{\mathbf{R}}
\def\r{\mathbf{r}}

\def\k{\mathbf{k}}

\begin{document}

\title{Model Hamiltonian for strongly correlated systems: Systematic, self-consistent, and unique construction}

\author{Ryan Requist}
\affiliation{Max Planck Institute of Microstructure Physics, Weinberg 2, 06120, Halle, Germany}
\author{E. K. U. Gross}
\affiliation{Max Planck Institute of Microstructure Physics, Weinberg 2, 06120, Halle, Germany}
\affiliation{
Fritz Haber Center for Molecular Dynamics, Institute of Chemistry, The Hebrew University of Jerusalem, Jerusalem 91904 Israel
}

\date{\today}

\begin{abstract}
An interacting lattice model describing the subspace spanned by a set of strongly correlated bands is rigorously coupled to density functional theory to enable {\it ab initio} calculations of geometric and topological material properties.  The strongly correlated subspace is identified from the occupation number band structure as opposed to a mean-field energy band structure. The self-consistent solution of the many-body model Hamiltonian and a generalized Kohn-Sham equation exactly incorporates momentum-dependent and crystal-symmetric correlations into electronic structure calculations in a way that does not rely on a separation of energy scales.  Calculations for a multiorbital Hubbard model demonstrate that the theory accurately reproduces the many-body macroscopic polarization.  
\end{abstract}

\maketitle

\section{\label{sec:introduction} Introduction}

Predicting electronic properties of strongly correlated materials is an enduring 
challenge in condensed matter physics.  Homogeneous electron gas based semilocal density functional approximations do not capture the strong correlations between electrons hopping within a manifold of localized states, as present in Mott insulators \cite{mott1949,mott1964,brandow1977,imada1998}, cuprate superconductors \cite{anderson1987a,monthoux1991,dagotto1994,lee2006,schrieffer2007} and heavy-fermion compounds \cite{stewart1984,freeman1987,fuggle1988,grewe1991,hewson1997}.  While progress has been made, there remains a need for {\it ab initio} computational methods capable of accurately predicting the emergent phenomena, phase diagrams and sensitive dependence on external parameters in such systems.

Density functional theory (DFT) calculations of correlated solids may encounter two types of deficiencies.  First, a calculation may deliver a qualitatively incorrect density or total energy.  By virtue of the Hohenberg-Kohn theorem \cite{hohenberg1964}, these two quantities should be correct if the exchange-correlation functional is accurate.  The inability of semilocal approximations to correctly predict orbital ordering in some compounds, e.g.,~KCuF$_3$ \cite{liechtenstein1995,towler1995,pavarini2008,leonov2010}, LaMnO$_3$ \cite{elfimov1999,yin2006,pavarini2010,leonov2010}, and KCrF$_3$ \cite{autieri2014,novoselov2016}, implies that not only the densities and energies but also the structures are incorrect, since they lack the associated symmetry-lowering Jahn-Teller distortion.  Similarly, spin-DFT \cite{vonbarth1972,gunnarsson1976} calculations of the spin state or pressure-induced spin state crossover in transition metal atoms in oxides \cite{cohen1997} may fail qualitatively, or suffer from large uncertainties, as seen for Fe in MgO \cite{korotin1994,tsuchiya2006a} and MgSiO$_3$ \cite{zhang2006,umemoto2008,hsu2010}. 

A second type of deficiency occurs when the Kohn-Sham system \cite{kohn1965}, an auxiliary noninteracting system that reproduces the density of the interacting system, does not provide a qualitatively correct reference state for subsequent higher-level calculations.  In the absence of spin-symmetry breaking, Kohn-Sham band structures are metallic for Mott insulators and as such do not provide an appropriate reference state for calculations of the macroscopic polarization and related quantities.  An important unsolved problem is to find a way to perform accurate {\it ab initio} calculations of topological invariants in strongly correlated systems.  Topological invariants were originally formulated in terms of the Bloch states of a mean-field band structure \cite{thouless1982,haldane1988,kane2005,fu2007,bernevig2006,moore2007,roy2009}, almost invariably the Kohn-Sham band structure, an approach that may give incorrect results in strongly correlated systems.  Although interacting topological invariants can be rigorously defined in terms of the Berry curvature of the correlated many-body wavefunction \cite{thouless1983,niu1984}, those formulas have not been applied in {\it ab initio} calculations of real materials due to the formidable difficulty of approximating the correlated wavefunction of an infinite solid.  The mean-field-based geometric phase formula for the macroscopic polarization \cite{king-smith1993,resta1994} has been similarly generalized to interacting systems \cite{ortiz1994}, but the latter formula has not been applied to real materials for the same reason. 

To study the phases and physical properties of strongly correlated materials, one typically introduces an effective Hamiltonian defined on a lattice and comprising a small but relevant subset of low-energy degrees of freedom.  Thus, the single narrow band in the Hubbard model \cite{hubbard1963,kanamori1963,gutzwiller1963,anderson1959} represents the Mott insulator-metal transition in transition metal oxides, while the coexistence of localized and itinerant electrons in the periodic Kondo and Anderson models \cite{doniach1977,tsunetsugu1997,hewson1997} embodies the key physics of heavy fermion compounds.  At this level of theory, the interaction between high-energy and low-energy degrees of freedom is {\it one-way}: the coupling to high-energy states is assumed to renormalize the parameters of the low-energy Hamiltonian, but the effects of many-body quantum fluctuations within the correlated low-energy sector on the high-energy degrees of freedom are neglected.  Effective low-energy models can be rigorously derived by a procedure called downfolding \cite{loewdin1951,loewdin1962,brandow1979}. 

The downsides of model Hamiltonian approaches are ambiguities in the choice of the model and uncertainties in the model parameters.  Quantitative results depend on the precise values of the model parameters, which are often estimated from {\it ab initio} calculations or fixed empirically by comparison with experiment.  Empirical approaches have limited predictive power, and often there are more relevant parameters than could conceivably be determined by fitting to experiment.  DFT has long been used to guide the choice of the relevant orbital, spin, and lattice degrees of freedom to be included in a model Hamiltonian and to estimate their mutual interactions.  The one-body hopping and on-site potential terms of approximate tight-binding models are routinely derived by fitting \cite{papaconstantopoulos1986,horsfield2000,papaconstantopoulos2003} Kohn-Sham band structures with Slater-Koster parameters \cite{slater1954} or by transforming the Kohn-Sham Hamiltonian from the basis of Bloch states to a basis of localized (Wannier) functions \cite{andersen1984,mattheiss1989,mcmahan1990,marzari1997,ku2002}.  

The definition of two-body interaction parameters, on the other hand, is fraught with difficulty.  The largest Coulomb matrix element is the Hubbard interaction $U$ between two electrons occupying the same atomic-like (e.g.~$d$ or $f$) orbital.  The definition $U=E(d^{n+1}) + E(d^{n-1}) - 2E(d^n)$ \cite{vanvleck1953} in terms of atomic configurations with different numbers of localized $d$ electrons self-consistently screened by conduction electrons \cite{herring1966} agrees well with photoemission spectra for very localized orbitals, such as the $4f$ orbitals in rare-earth metals \cite{herbst1978}, but still needs to be rigorously connected to an interaction term $U \hat{n}_{i\u} \hat{n}_{i\d}$ in a model Hamiltonian.  If one associates a localized Wannier function $w_i(\r)=w(\r-\R_i)$ to the lattice site $i$ of a model Hamiltonian, then it is natural to define $U$ as the on-site matrix element of the Coulomb interaction, $U=\langle w_i w_i | V_{ee}|w_i w_i\rangle$ \cite{anderson1959,anderson1961,hubbard1963}.  However, this gives unrealistically large values, since the bare Coulomb interaction $V_{ee}(\r,\r')=e^2/|\r-\r'|$ should be screened \cite{anisimov1997a,springer1998,kotani2000,biermann2003}.  Interatomic and interorbital Coulomb matrix elements are usually not negligible and, if not included in the model Hamiltonian, contribute to further renormalizing the value of $U$ \cite{gunnarsson1989b,gunnarsson1990}.  Screening is accounted for in the {\it constrained random phase approximation} \cite{aryasetiawan2004,miyake2009,vaugier2012}, where Coulomb matrix elements are defined as $U_{ijkl}=\langle w_i w_j | \epsilon^{-1}V_{ee}|w_k w_l\rangle$.  The dielectric function $\epsilon$ is calculated from an {\it ab initio} irreducible electronic polarizability in which the contribution from a low-energy correlated subspace has been subtracted out to avoid double-counting screening channels that are already present in the many-body model.  Since the width and shape of conventional Wannier functions depend on the gauge of the Bloch functions from which they are constructed, the $U_{ijkl}$ determined in this way are dependent on the specific gauge choices that are made. 

The {\it constrained occupation method} in DFT provides a seamless estimate of effective interaction parameters by relating them to how the energy changes in response to changes in the occupation or magnetic moment of a local orbital \cite{dederichs1984,min1986,mcmahan1988,gunnarsson1989a,hybertsen1989}.  Although the effective $U$ is thus calculated self-consistently in the presence of all screening channels and with only the information contained in the exchange-correlation functional, it still depends on the arbitrary definition of the local orbital whose occupation is to be constrained.  All existing approaches to the calculation of model parameters suffer from this fundamental nonuniqueness.  Even the one-body terms evaluated in Wannier-based tight-binding approaches are not unambiguously defined, as the hopping amplitudes depend implicitly on the many-body configuration of the relevant local orbitals \cite{gunnarsson1988b,gunnarsson1989a} and should be renormalized by Coulomb interactions \cite{gunnarsson1989b}. 

A general procedure for deriving effective low-energy models is to integrate out the high-energy degrees of freedom.  The Hamiltonian for the high-energy sector and the interaction terms that couple the high-energy and low-energy sectors are ``downfolded'' \cite{loewdin1951,loewdin1962,brandow1979} into an effective operator that acts on the low-energy Hilbert space, thus defining an effective low-energy model with renormalized Hamiltonian parameters \cite{gunnarsson1990,andersen1995}. This strategy has developed into a widely used {\it ab initio downfolding method} \cite{lichtenstein1998,aryasetiawan2004,solovyev2005,solovyev2008,miyake2008,aryasetiawan2009,miyake2009,imada2010,vaugier2012,casula2012,sasioglu2013,hirayama2013,hirayama2017,nilsson2018}.  After defining an energy window for the construction of localized Wannier functions and choosing a correlated subspace spanned by Wannier functions of specific orbital character, the frequency-dependent interaction parameters of an effective low-energy model are calculated by applying the constrained random phase approximation to the disentangled band structure.  The effective low-energy model depends on the number and character of the Wannier orbitals in the correlated subspace, as well as the energy window and gauge choices used in constructing the Wannier functions \cite{vaugier2012,miyake2009}.

Downfolding methods are justified when there is a separation of energy scales.  In {\it ab initio} downfolding methods, the separation into low- and high-energy subspaces, as well as the definition of the Wannier orbital basis for the many-body model, are based on the mean-field Kohn-Sham band structure.  The low-energy localized Wannier orbitals are chosen as the degrees of freedom to be correlated at a higher level of theory by solving the many-body model \cite{miyake2009,vaugier2012,hirayama2017}.  Another approach to downfolding is to start from an {\it ab initio} quantum Monte Carlo calculation and use a fitting procedure to determine the effective Hamiltonian that best reproduces the two-body reduced density matrix in a low-energy sector \cite{changlani2015}.

Many materials of current interest have complex multiband character with several competing interactions, involving charge, spin, orbital and lattice degrees of freedom, making it hard to arrive at a unique model Hamiltonian.  In several cases, multiple different models have been introduced to describe the same material property; for example, one-band \cite{anderson1987a,zhang1988} and three-band \cite{emery1987,varma1987,abrahams1987,hirsch1987} extended Hubbard models for high $T_c$ cuprate superconductors with parameters estimated early on from DFT band structure calculations \cite{mattheiss1987,yu1987,mcmahan1988,hybertsen1989,mattheiss1989,pickett1989,mcmahan1990,andersen1995};  a Kane-Mele-type model \cite{shitade2009} and a Kitaev-Heisenberg model \cite{jackeli2009,chaloupka2010} with a raft of additional interactions \cite{kimchi2011,albuquerque2011,choi2012,singh2012,ye2012,chaloupka2013,rau2014,bhattacharjee2012,sizyuk2014,winter2016,laubach2017} for sodium iridate (Na$_2$IrO$_3$) and its zigzag spin ordering \cite{liu2011,choi2012,ye2012}, which has also been studied with first-principles calculations \cite{foyevtsova2013,yamaji2014,katukuri2014,hu2015}; 
one-band \cite{perfetti2005,perfetti2006,sipos2008,lahoud2014} and multiband \cite{qiao2017} Hubbard models, possibly with additional interlayer coupling \cite{bovet2003,ritschel2015}, spin-orbit coupling \cite{rossnagel2006}, disorder \cite{disalvo1975,mutka1981,zwick1998,lahoud2014} and Hubbard-Holstein renormalization \cite{cho2015} effects, for the Mott \cite{tosatti1976,fazekas1979} and superconducting \cite{morosan2006,sipos2008,ang2013} phases in the charge density wave state \cite{wilson1975,scruby1975} of the transition metal dichalcogenide 1$T$-TaS$_2$.  The challenges one faces in defining a unique model Hamiltonian make it difficult to reach agreement on the underlying physical explanation for the observed phases and emergent phenomena, particularly for cuprate superconductors, where various issues have been debated for decades \cite{multiple2006}.  Does every material have a model Hamiltonian that is unique in some well-defined sense for a chosen subset of variables?  

While DFT is the standard framework for itinerant, nearly free-electron-like states and model Hamiltonians are widely used for strongly correlated localized states, both approaches have limitations when local and itinerant electrons coexist.  The advantages of treating localized and itinerant electrons differently was appreciated long ago in the works of Anderson, Hubbard, Kanamori, Gutzwiller and others \cite{anderson1959,anderson1961,hubbard1963,kanamori1963,gutzwiller1963}.  Several methods that combine model Hamiltonians with DFT have since been developed.  Most approaches involve a mapping to an auxiliary Anderson impurity model, i.e.~they single out an atomic-like orbital with strong on-site interactions and treat it as an impurity that hybridizes with a band of noninteracting electrons representing the remaining delocalized degrees of freedom of the solid.  In DFT+dynamical mean field theory (DFT+DMFT) \cite{anisimov1997b,lichtenstein1998,kotliar2006,held2006}, the frequency-dependent hybridization function of the Anderson impurity model is determined by requiring self-consistency between the local lattice Green's function calculated within DFT and the impurity Green's function of the Anderson model.  In the DFT+numerical renormalization group (DFT+NRG) approach \cite{lucignano2009,baruselli2012,baruselli2013,requist2014,baruselli2015}, frequency-independent impurity model parameters are determined by equating the mean-field scattering phase shifts of the Anderson model to the scattering phase shifts calculated within DFT.  In density matrix embedding theory (DMET), the Schmidt decomposition between a few localized states and the rest of the system leads to an effective Anderson model in which the frequency-independent coupling to bath states is determined by imposing self-consistency on the local reduced density matrix \cite{knizia2012}.  In site occupation embedding theory \cite{fromager2015,senjean2017}, one solves self-consistently for the site occupation numbers of a lattice model, with Hubbard interactions turned on only in a fragment consisting of a few sites, in the presence of an embedding potential derived from a bath correlation energy functional.

By limiting correlations to the impurity model subspace, DFT+DMFT and all impurity-based embedding approaches explicitly break lattice translational symmetry in the correlated part of the problem.  As a result, they do not provide information on nonlocal momentum-dependent correlations.  This will cause inaccuracies in the calculation of geometric and topological quantities, which are specifically related to the $\k$-dependence of the single-particle Bloch functions in the mean-field case and the total quasimomentum-dependence of the correlated wavefunction in the interacting case.  Geometric and topological properties are therefore sensitive to the momentum-dependence of two-body correlations in strongly correlated systems.

In this article, we propose an {\it ab initio} theory that rigorously couples DFT to a unique many-body lattice model.  The strongly correlated subspace described by the lattice model is chosen by selecting a subset of natural occupation number bands ($\k$-dependent eigenvalues of the one-body reduced density matrix) that are isolated from all others.  Since natural occupation number bands are intrinsic variables directly calculable from the many-body wave function without the introduction of auxiliary mean-field quantities, they may provide a more accurate partitioning into strongly  and weakly correlated subspaces than a partitioning based on mean-field energy bands.  The one-body reduced density matrix is invariant with respect to the symmetry group of the crystal, and therefore its eigenfunctions, called natural Bloch orbitals, transform exactly as mean-field Bloch functions do and can be labeled by a wavevector $\k$ that is an element of the true Brillouin zone of the crystal.  The many-body model Hamiltonian expressed in the subspace of strongly correlated natural Bloch orbitals therefore preserves lattice translational symmetry and all other symmetries of the crystal.  Since the Hamiltonian model parameters are frequency independent, the many-body problem can be solved more efficiently than models with frequency-dependent parameters, for which a Lagrangian formalism is necessary.  The lack of frequency dependence is an exact feature of the theory and does not imply that dynamical correlations are neglected. 

Our theory provides a practical way to make {\it ab initio} calculations of geometric and topological properties in strongly correlated systems.  It has recently been shown that the natural Bloch orbitals, natural occupation numbers and their conjugate phases, which combine to form a set of natural orbital geometric phases, contain most of the information about the effects of many-body correlations on geometric and topological quantities in the Rice-Mele-Hubbard model \cite{requist2018}.  Since these variables are included in our theory through the self-consistent solution of the many-body lattice model and a generalized Kohn-Sham equation, we expect to obtain accurate results for geometric and topological quantities in real systems.  These quantities can be problematic in standard DFT.  For example, the King-Smith--Vanderbilt formula \cite{king-smith1993} for the macroscopic polarization is undefined for Mott insulators and other systems for which the Kohn-Sham system is metallic.  Calculating topological invariants in terms of mean-field Bloch states is similarly problematic, since it is only by neglecting the interaction-induced broadening of the mean-field band structure that one obtains a quantized result.  In contrast, the natural occupation numbers form exact, unbroadened bands irrespective of the interaction strength.  Hence, topological invariants calculated in terms of the natural Bloch orbitals are precisely quantized \cite{requist2018}.  Quantum Monte Carlo methods have been used to evaluate many-body topological invariants in correlated model systems \cite{hohenadler2013}.  Although the application of such many-body methods to real systems with all electronic degrees of freedom is computationally prohibitive, they could be used instead of exact diagonalization to solve the model Hamiltonian in our theory, which self-consistently retains all electronic degrees of freedom in a generalized DFT framework.

Our main results from numerical calculations for a two-orbital Hubbard model are (i) a demonstration that two strongly correlated natural occupation number bands split off from the weakly correlated bands as the on-site interaction $U$ increases and (ii) a demonstration that our theory accurately predicts the many-body macroscopic polarization in strongly, weakly and intermediately correlated regimes of the model. 

The article is organized as follows.  In Sec.~\ref{sec:theory}, we present the fundamentals of the theory.  In Sec.~\ref{sec:application}, we define the two-orbital Hubbard model for which all calculations are made.  In Sec.~\ref{sec:bandstructure}, we calculate the exact natural occupation number band structure and demonstrate the partitioning into weakly and strongly correlated bands.  In Sec.~\ref{sec:model}, we evaluate the model Hamiltonian for the strongly correlated bands and investigate its density dependence.  In Sec.~\ref{sec:polarization}, we calculate the many-body polarization and verify that it is given correctly in our theory.  We provide conclusions and an outlook on future challenges in Sec.~\ref{sec:outlook}.

\section{\label{sec:theory} Theory}

Our theory is based on the idea of partitioning the natural occupation number bands into weakly and strongly correlated subsets and treating the latter at a higher level of theory.  To define the natural occupation number bands, we first need the one-body reduced density matrix 
\begin{align}
\rho_{1,\s\s'}(\r,\r') = \mathrm{Tr}\big[\hat{\psi}_{\s'}^{\dag}(\r') \hat{\psi}_{\s}(\r)\, \hat{\rho}\big] {,}
\end{align}
where $\hat{\psi}_{\s}^{\dag}(\r)$ and $\hat{\psi}_{\s}(\r)$ are the electronic creation and annihilation operators and $\hat{\rho}=\sum_{N,\alpha} w_{N\alpha}|\Psi_{N\alpha}\rangle \langle\Psi_{N\alpha}|$ is the density matrix of the system, which at equilibrium is an ensemble of $N$-electron eigenstates with weights $w_{N\alpha}=\exp[-\beta(E_{N\alpha}-\mu N)]/\sum_{N\alpha} \exp[-\beta(E_{N\alpha}-\mu N)]$.  Our theory is formulated in the grand canonical ensemble for a system with temperature $\tau$ ($\beta=1/k_B\tau$) and chemical potential $\mu$.  Since $\hat{\rho}_1$ commutes with lattice translations $\hat{T}_{\mathbf{a}}$, its eigenfunctions, which we call {\it natural Bloch orbitals} $\phi_{n\k}(\r)$, obey the Bloch condition $\phi_{n\k}(\r+\mathbf{a}) = e^{i\k\cdot \mathbf{a}} \phi_{n\k}(\r)$ and can be labeled by a band index $n$, a wavevector $\k$, and other possible quantum numbers associated with the crystallographic space group.  The natural Bloch orbitals are spin-orbitals, or generally two-component spinors $\phi_{n\k}(\r) = \{\phi_{n\k\u}(\r), \phi_{n\k\d}(\r)\}$, determined by the eigenvalue equation
\begin{align}
\sum_{\s'}\int \rho_{1\s\s'}(\r,\r') \phi_{n\k\s'}(\r') d\r' = f_{n\k} \phi_{n\k\s}(\r) {.}
\label{eq:one-rdm}
\end{align}
The occupation numbers $f_{n\k}$ form bands in the Brillouin zone of the crystal \cite{requist2018}.  In our theory, these natural occupation number bands take the place of mean-field energy bands.  Bands whose occupation numbers differ significantly from 0 and 1, if present, will be called strongly correlated, and the remaining bands whose occupation numbers are close to 0 or 1 will be called weakly correlated.  The natural Bloch orbitals are correspondingly partitioned into weakly and strongly correlated sets $\mathcal{S}$ and $\mathcal{D}$.  In DFT, only a finite number of energy bands are occupied at $\tau=0$.  In contrast, there are generally an infinite number of nonvanishing natural occupation number bands even at $\tau=0$, due to many-body correlations.    If the $f_{n\k}$ are ordered in a nonincreasing sequence $f_{1\k},f_{2\k},\ldots$ for each $\k$, there will not generally exist a lower bound $b_{\k}>0$ such that $f_{n\k}\geq b_{\k} \;\forall n$; zero is an accumulation point of the spectrum. 

Since there have been no many-body calculations of the natural occupation number bands in real materials, the extent to which they can be unambiguously partitioned into weakly and strongly correlated subsets is presently unknown.  Our numerically exact results for a two-orbital Hubbard model (Fig.~\ref{fig:bands:series:U}) provide a concrete example of a system where two strongly correlated bands clearly split off from the remaining two weakly correlated bands as the Hubbard interaction is increased.  The distinction between weakly and strongly correlated bands is only an approximate notion, e.g.~if one occupation number band hovers around 0.98 and another around 0.99, it would hardly be possible to argue that the former is more correlated than the latter.  Moreover, if a band whose occupation numbers differ significantly from 0 and 1 in some region of the Brillouin zone has symmetry-enforced or accidental intersections with weakly correlated bands or is otherwise strongly ``entangled'' with them, it might be questionable to single it out for special treatment.  Nevertheless, our theory is formally exact (in the DFT sense of returning the exact equilibrium values of the functional variables) for any partitioning of the occupation number bands.  Specifically, if none of the bands are treated as strongly correlated, then our theory reduces to DFT (or current-DFT \cite{vignale1987}).  On the other hand, if all of the bands are treated as strongly correlated, then the model Hamiltonian is simply the full many-body Hamiltonian expressed in the basis of natural Bloch orbitals.  For real strongly correlated materials, where a few relevant bands would be treated at the full many-body level, we expect that it will be easier to find accurate functional approximations in our theory than in conventional DFT.
 
Even in cases where it is not possible to cleanly disentangle the bands, there may still be advantages to using the natural Bloch orbitals or natural Wannier functions \cite{requist2018}, as opposed to mean-field Bloch orbitals or mean-field Wannier functions, in selecting a subset of degrees of freedom to treat at a higher level of theory.  Natural Bloch orbitals are intrinsic variables of the system, being defined in terms of the one-body reduced density matrix, a quantity obtained by simply tracing out degrees of freedom---a {\it linear} operation---and may therefore provide a more suitable starting point than mean-field Bloch orbitals in strongly correlated systems.  

The next step is to introduce a generalized density functional theory in which the basic variables are the density $n(\r)$, the paramagnetic current density $\mathbf{j}_p(\r)$, and the strongly correlated natural Bloch orbitals $\phi_{d\k\s}(\r)$.  The paramagnetic current density is included to correctly account for the coupling to an artificial vector potential that will be introduced to evaluate the macroscopic polarization, yet we expect that the $\mathbf{j}_p(\r)$-dependence of the functionals can be neglected as a first approximation.  As we are working at temperature $\tau$ and chemical potential $\mu$, we define the following grand potential functional for a system of interacting electrons in the presence of scalar and vector potentials $v(\r)$ and $\mathbf{A}(\r)$ (a uniform $\mathbf{A}$ is sufficient for our purposes \footnote{A uniform vector potential $\mathbf{A}$ will be used to apply an artificial magnetic flux to the system, viewed as living on a torus, which is equivalent to imposing twisted boundary conditions \cite{kohn1964,thouless1983,niu1984}.}) \cite{mermin1965}:
\begin{align}
\Omega[n,\mathbf{j}_p,\phi_{d\k},f_{d\k},\rho_2^d] &= \int [v(\r)-\mu] n(\r) d\r \nn \\
&\quad+ \f{e}{c} \int \mathbf{A}(\r)\cdot \mathbf{j}_p(\r)d\r \nn \\
&\quad+ \f{e^2}{2mc^2} \int |\mathbf{A}(\r)|^2 n(\r) d\r \nn \\
&\quad+ F[n,\mathbf{j}_p,\phi_{d\k},f_{d\k},\rho_2^d] {,}
\label{eq:Omega}
\end{align}
where the universal functional $F$ is defined by the following constrained search \cite{levy1979} over density matrices $\rho$ that yield the set of variables $X=(n,\mathbf{j}_p,\phi_{d\k},f_{d\k},\rho_2^d)$:
\begin{align}
F[n,\mathbf{j}_p,\phi_{d\k},f_{d\k},\rho_2^d] = \min_{\rho\rightarrow X} \mathrm{Tr}[(\hat{T}+\hat{V}_{ee}-\tau \hat{S})\hat{\rho}] {,}
\end{align}
where $\hat{T}$ is the kinetic energy, $\hat{V}_{ee}$ is the electron-electron interaction, and $\hat{S} = -k_B \ln \hat{\rho}$ is the entropy operator.  The functionals $\Omega$ and $F$ additionally depend on the natural occupation numbers $f_{d\k}$ and the two-body reduced density matrix $\rho_2^d$ in the strongly correlated $\mathcal{D}$ subspace.  The matrix elements of $\rho_2^d$ are 
\begin{align}
(\rho^d_2)_{121'2'} &= \f{1}{2} \mathrm{Tr}(c_{2'}^{\dag} c_{1'}^{\dag} c_1 c_2 \, \hat{\rho}) {,}
\label{eq:Gammad}
\end{align}
where $1=(d_1\k_1)$ and $c_1^{\dag}$ is the creation operator for an electron in the natural Bloch orbital state $\phi_{d_1\k_1} \in \mathcal{D}$.  The grand potential can be further decomposed by defining a kinetic-energy-entropy functional for noninteracting electrons (with $\hat{S}_s=-k_B \ln \hat{\rho}_s$)
\begin{align}
K_s[n,\mathbf{j}_p,\phi_{d\k},f_{d\k}] &= \min_{\rho_s \rightarrow (n,\mathbf{j}_p,\phi_{d\k},f_{d\k})} \mathrm{Tr}[(\hat{T}-\tau\hat{S}_s)\hat{\rho}_s] 
\label{eq:K}
\end{align}
and a Hartree-exchange-correlation grand potential 
\begin{align}
\Omega_{hxc}[n,\mathbf{j}_p,\phi_{d\k},f_{d\k},\rho_2^d] &= F[n,\mathbf{j}_p,\phi_{d\k},f_{d\k},\rho_2^d] \nn \\
&\quad- K_s[n,\mathbf{j}_p,\phi_{d\k},f_{d\k}] {.}
\label{eq:Omega:hxc}
\end{align}
The kinetic-energy-entropy functional $K_s[n,\mathbf{j}_p,\phi_{d\k},f_{d\k}]$, defined by a constrained search over ensembles of Slater determinants $\rho_s$, is  intermediate between the corresponding functionals in DFT \cite{pittalis2011} and reduced density matrix functional theory \cite{baldsiefen2015}.  Since $K_s[n,\mathbf{j}_p,\phi_{d\k},f_{d\k}]$ accounts for the fractional occupation numbers of the strongly correlated natural Bloch orbitals, we expect it to provide a better approximation to the true kinetic-energy-entropy than the DFT functional.  $\Omega$ and $F$ are defined on the domain of $(n,\mathbf{j}_p,\phi_{d\k},f_{d\k},\rho_2^d)$ that can be obtained from a fermionic density matrix, which we denote as the ensemble representable (ER) domain in analogy to the ensemble $N$ representable domain for fixed particle number $N$ \cite{coleman1963}.  Using the quantum generalization of Gibb's variational principle for ensembles \cite{mermin1965}, it is straightforward to prove the following variational principle.

{\it Theorem}.--- The grand potential functional satisfies $\Omega[n,\mathbf{j}_p,\phi_{d\k},f_{d\k},\rho_2^d]>\Omega_0$ for any $(n,\mathbf{j}_p,\phi_{d\k},f_{d\k},\rho_2^d)$ that are not equal to the correct equilibrium variables $(n_0,\mathbf{j}_{p0},\phi_{d\k0},f_{d\k0},\rho^d_{20})$ yielding the potential $\Omega_0$.

The density and paramagnetic current density can be expressed in terms of the natural Bloch orbitals as
\begin{align}
n(\r) &= \sum_{n\k\s} f_{n\k} |\phi_{n\k\s}(\r)|^2 \nn \\
\mathbf{j}_p(\r) &= \f{\hbar}{m} \mathrm{Im} \sum_{n\k\s} f_{n\k} \phi_{n\k\s}^*(\r)\nabla \phi_{n\k\s}(\r)  {.} \label{eq:n:all}
\end{align}
Our strategy is now to postulate that a semilocal density functional \cite{kohn1965,vonbarth1972,perdew1996} provides a sufficiently accurate description of the weakly correlated bands in the sense that their contribution to the density in Eq.~(\ref{eq:n:all}) can be well-approximated by a sum $\sum_{\s,n\k\in\mathcal{S}} g_{n\k}|\chi_{n\k\s}(\r)|^2$ over KS-like Bloch orbitals $\chi_{n\k\s}(\r)$ that are eigenstates of a noninteracting Hamiltonian with a scalar multiplicative potential $v_s(\r)$.  The thermal occupation numbers in this sum, $g_{n\k}=(1+\exp[\beta(\epsilon_{n\k}-\mu)])^{-1}$, follow the Fermi-Dirac distribution and therefore manifest thermal fluctuations but not quantum fluctuations.  At the same time, we retain the fractional occupation numbers $f_{d\k}$ of the strongly correlated natural Bloch orbitals $\phi_{d\k}(\r)$ as variational parameters, since they do exhibit significant quantum fluctuations.  

The grand potential is minimized in an iterative fashion.  First, for fixed $(f_{d\k},\rho_2^d)$, $\Omega$ is minimized with respect to $n(\r)$, $\mathbf{j}_p(\r)$, and $\phi_{d\k\s}(\r)$ by finding the self-consistent solution of a generalized Kohn-Sham equation (see Appendix A) 
\begin{align}
\hat{h}^{\mathit{eff}} |\psi_{b\k} \rangle = \epsilon_{b\k} |\psi_{b\k}\rangle
\label{eq:GKS}
\end{align}
with the Hamiltonian ($e$ is the absolute value of charge)
\begin{align}
\hat{h}^{\mathit{eff}} &= \f{1}{2m} \big(\hat{\mathbf{p}} + \f{e}{c} \mathbf{A}_s(\hat{\r})\big)^2 + v_s(\hat{\r})  \nn \\
&\quad+ \sum_{d\k d'\k'} w_{d\k,d'\k'} | \phi_{d\k} \rangle \langle \phi_{d'\k'} | {,} \label{eq:h}
\end{align} 
where $\mathbf{A}_s(\r)=\mathbf{A}(\r)+\mathbf{A}_{xc}(\r)$ and $v_s(\r) = v(\r) + v_{hxc}(\r) + e^2(|\mathbf{A}(\r)|^2 - |\mathbf{A}_s(\r)|^2)/2mc^2$, similar to current-density functional theory \cite{vignale1987}.  The set of eigenfunctions $\{\psi_{b\k}\}$ of $\hat{h}^{\mathit{eff}}$ contains the strongly correlated orbitals $\phi_{d\k}$ as well as the Kohn-Sham-like orbitals $\chi_{n\k}$.  The latter span the same space as the weakly correlated natural Bloch orbitals $\phi_{n\k}$.  These orbitals, together with the strongly correlated natural occupation numbers $f_{d\k}$, are used to evaluate the density and paramagnetic current density  
\begin{align}
n(\r) &= \sum_{\s,n\k} g_{n\k} |\chi_{n\k\s}(\r)|^2 + \sum_{\s,d\k\in \mathcal{D}} f_{d\k} |\phi_{d\k\s}(\r)|^2 \nn \\
\mathbf{j}_p(\r) &= \f{\hbar}{m}\mathrm{Im} \Big[ \sum_{\s,n\k} g_{n\k} \chi_{n\k\s}^*(\r)\nabla \chi_{n\k\s}(\r) \Big] \nn \\
&\quad+ \f{\hbar}{m}\mathrm{Im} \Big[\sum_{\s,d\k\in \mathcal{D}} f_{d\k} \phi_{d\k\s}^*(\r)\nabla \phi_{d\k\s}(\r) \Big] {.} \label{eq:n}
\end{align}

Second, $\Omega$ is minimized with respect to $f_{d\k}$ and $\rho_2^d$ on the ER domain for fixed $(n,\mathbf{j}_p,\phi_{d\k})$.  Instead of minimizing $\Omega$ directly, we minimize the grand potential of an auxiliary system describing only the strongly correlated subspace and constructed to have the same $f_{d\k}$ and $\rho_2^d$ at its minimum.  As a first step to defining the auxiliary grand potential, the full density matrix is expanded as
\begin{align}
\hat{\rho} = \sum_i a_i \hat{\mu}_i + \sum_i b_i \hat{\nu}_i + \sum_i c_i \hat{\xi}_i + \sum_i d_i \hat{o}_i {,}
\label{eq:rho:expansion}
\end{align}
where $\hat{\mu}_i$, $\hat{\nu}_i$, $\hat{\xi}_i$ and $\hat{o}_i$ are a complete basis of Hermitian operators that are mutually orthonormal with respect to the Hilbert-Schmidt inner product $\langle \hat{A},\hat{B}\rangle = \mathrm{Tr}(\hat{A}^{\dag} \hat{B})$.  The $\hat{\mu}_i$ form a complete basis of one-body operators in the $\mathcal{D}$ subspace, the $\hat{\nu}_i$ form a complete basis for all two-body operators in the $\mathcal{D}$ subspace that are linearly independent of all $\hat{\mu}_i$, the $\hat{\xi}_i$ form a complementary basis of one- and two-body operators that are linearly independent of all $\hat{\mu}_i$ and $\hat{\nu}_i$, and $\hat{o}_i$ span all remaining three-body, four-body, $\ldots$ operators \cite{requist2014b}; for details see Appendix B.  Since $\hat{\mu}_i$, $\hat{\nu}_i$ and $\hat{\xi}_i$ are a complete one- and two-body basis, we can expand the full two-body reduced density matrix as
\begin{align}
\hat{\rho}_2 = \sum_i a_i \hat{\mu}_i + \sum_i b_i \hat{\nu}_i + \sum_i c_i \hat{\xi}_i
\end{align}
and the reduced density matrix in the $\mathcal{D}$ subspace as
\begin{align}
\hat{\rho}_2^d = \sum_i a_i \hat{\mu}_i + \sum_i b_i \hat{\nu}_i {.}
\end{align}
The one-body reduced density matrix in the $\mathcal{D}$ subspace, which contains the information about $f_{d\k}$, can be written as $\hat{\rho}_1^d = \sum_i a_i \hat{\mu}_i$.  Finally, the Hamiltonian, which is assumed to contain only one-body and two-body terms, can be expanded as
\begin{align}
\hat{H} - \mu \hat{N} = \sum_i V_i \hat{\mu}_i + \sum_i U_i \hat{\nu}_i+ \sum_i W_i \hat{\xi}_i {.}  
\label{eq:H:op}
\end{align}
Next, we define the auxiliary grand potential functional
\begin{align}
\Omega^{aux}[a_i,b_i|n,\mathbf{j}_p,\phi_{d\k}] &= \sum_i V_i a_i + \sum_i U_i b_i \nn \\
&\quad+ \mathcal{F}[a_i,b_i|n,\mathbf{j}_p,\phi_{d\k}] {,}
\end{align}
where 
\begin{align}
\mathcal{F}[a_i,b_i|n,\mathbf{j}_p,\phi_{d\k}] = \min_{\rho\rightarrow a_i,b_i,n,\mathbf{j}_p,\phi_{d\k}} \mathrm{Tr}\Big[\Big(\sum_i W_i \hat{\xi}_i - \tau \hat{S}\Big) \hat{\rho}\Big]
\end{align}
is a universal functional that does not depend on $\{V_i,U_i\}$.

Now we use a reductio ad absurdum argument to prove that $\rho_1^d$ and $\rho_2^d$ uniquely determine $\{V_i,U_i\}$ for fixed $\{W_i\}$.  Consider two Hamiltonians $\hat{H}$ and $\hat{H}'$ with $\{V_i,U_i\}$ and $\{V_i',U_i'\}$ that are different.  The corresponding equilibrium density matrices are denoted $\rho$ and $\rho'$.  Suppose that $\rho$ and $\rho'$ yield the same $\rho_1^d$ and $\rho_2^d$, i.e.~have the same $\{a_i,b_i\}$.  Then, according to the variational principle for the grand potential \cite{mermin1965}, we have 
\begin{align}
\mathrm{Tr}[(\hat{H}-\mu\hat{N}-\tau\hat{S})\hat{\rho}] &< \mathrm{Tr}[(\hat{H}-\mu\hat{N}-\tau\hat{S}')\hat{\rho}'] \nn \\ 
\mathrm{Tr}[(\hat{H}'-\mu\hat{N}-\tau\hat{S}')\hat{\rho}'] &< \mathrm{Tr}[(\hat{H}'-\mu\hat{N}-\tau\hat{S})\hat{\rho}] {,}
\end{align}
which leads to
\begin{align}
\mathrm{Tr}\Big[\sum_i (V_i-V_i') \hat{\mu}_i \sum_j (a_j-a_j') \hat{\mu}_j\Big] \nn \\
+\mathrm{Tr}\Big[\sum_i (U_i-U_i') \hat{\nu}_i \sum_j (b_j-b_j') \hat{\nu}_j\Big] &< 0 \nn \\
\sum_i (V_i-V_i')(a_i-a_i') + \sum_i (U_i-U_i')(b_i-b_i') &< 0 {.}
\end{align}
Since having $a_i = a_i'$ and $b_i=b_i'$ for all $i$ would give a contradiction, we find that if $V_i \neq V_i'$ or $U_i \neq U_i'$ for some $i$, then $(\rho_1^d,\rho_2^d) \neq (\rho_1^{d\prime},\rho_2^{d\prime})$; hence, there is a unique mapping $\{a_i,b_i\} \rightarrow \{V_i,U_i\}$ \footnote{For $\tau=0$ and particle number $N$ we cannot prove $\{a_i,b_i\} \rightarrow \{V_i,U_i\}$, but we can still define energy functionals and a model Hamiltonian, though the latter may have a degree of nonuniqueness associated with conserved quantities \cite{capelle2001}.}.  $\Omega^{aux}$ and $\mathcal{F}$ are defined on the ER domain, and $\Omega^{aux}$ satisfies the variational principle $\Omega^{aux}[a_i,b_i|n,\mathbf{j}_p,\phi_{d\k}]> \Omega_0$ for $(a_i,b_i|n,\mathbf{j}_p,\phi_{d\k}) \neq (a_{i0},b_{i0}|n_0,\mathbf{j}_{p0},\phi_{d\k0})$.

A Kohn-Sham-type construction will be used to derive the model Hamiltonian for the auxiliary system.  We first define the functional
\begin{align}
\Lambda[a_i,b_i|n,\mathbf{j}_p,\phi_{d\k}] &= \Omega^{aux}[a_i,b_i|n,\mathbf{j}_p,\phi_{d\k}] \nn \\
&\quad+ \sum_n \lambda_n g_n[a_i,b_i|n,\mathbf{j}_p,\phi_{d\k}] {,}
\label{eq:Lambda:model}
\end{align}
where $\lambda_n$ are Karush-Kuhn-Tucker multipliers \cite{kuhn1951} that impose all ER constraints $g_n\leq 0$ on $\{a_i,b_i\}$.  In addition to the usual stationary conditions, the minimum $\{a_{i0},b_{i0}\}$ must satisfy the following necessary conditions for each ER constraint that is an inequality rather than a strict equality: (i) the complementary slackness condition $\lambda_n g_n[a_{i0},b_{i0}]=0$ and (ii) the feasibility conditions $g_n[a_{i0},b_{i0}]\leq 0$ and $\lambda_n\geq 0$ \cite{kuhn1951}. The stationary conditions $\partial \Lambda/\partial a_i=0$ and $\partial \Lambda/\partial b_i=0$ yield 
\begin{align}
V_i + \f{\partial \mathcal{F}}{\partial a_i} + \sum_n \lambda_n \f{\partial g_n}{\partial a_i} &= 0 \nn \\
U_i + \f{\partial \mathcal{F}}{\partial b_i} + \sum_n \lambda_n \f{\partial g_n}{\partial b_i} &= 0 {.} \label{eq:stat:conds}
\end{align}
Setting $W_i=0$, we define the functional
\begin{align}
\Lambda^d[a_i,b_i] &= \Omega^d[a_i,b_i] + \sum_n \lambda_{n}^d g_{n}^d[a_i,b_i] {,}
\end{align}
where $g_{n}^d[a_i,b_i]\leq 0$ are ER constraints and 
\begin{align}
\Omega^d[a_i,b_i] &= \sum_i V_{i}^{model} a_i + \sum_i U_{i}^{model} b_i + \mathcal{F}^d[a_i,b_i]
\end{align} 
with 
\begin{align}
\mathcal{F}^d[a_i,b_i] = \min_{\rho^d\rightarrow (a_i,b_i)} \mathrm{Tr}[- \tau \hat{S}^d \hat{\rho}^d] {.}
\end{align}
Here, $\hat{S}^d=-k_B\ln \hat{\rho}^d$ and $\hat{\rho}^d=\sum_{N\alpha} w_{N\alpha}^d |\Phi_{N\alpha}\rangle\langle \Phi_{N\alpha}|$ is a density matrix comprising many-body states built up exclusively from natural Bloch orbitals in the $\mathcal{D}$ subspace, i.e.~$|\Phi_{N\alpha}\rangle=\sum_D A_{N\alpha,D} c_{d_1\k_1}^{\dag} \ldots c_{d_N\k_N}^{\dag}|0\rangle$, where $D=(d_1\k_1,\ldots,d_N\k_N)$.  The stationary conditions for $\Lambda^d$ are the same as those in Eq.~(\ref{eq:stat:conds}) if we set 
\begin{align}
V_{i}^{model} &= V_i + \f{\partial (\mathcal{F}-\mathcal{F}^d)}{\partial a_i} + \sum_n \Big( \lambda_n \f{\partial g_n}{\partial a_i} - \lambda_{n}^d\f{\partial g_{n}^d}{\partial a_i} \Big) \nn \\ 
U_{i}^{model} &= U_i + \f{\partial (\mathcal{F}-\mathcal{F}^d)}{\partial b_i} + \sum_n \Big( \lambda_n \f{\partial g_n}{\partial b_i} - \lambda_{n}^d\f{\partial g_{n}^d}{\partial b_i} \Big) {.}
\end{align}  
This defines a model Hamiltonian
\begin{align}
\hat{H}^{model} - \mu \hat{N}^d = \sum_i V_{i}^{model} \hat{\mu}_i + \sum_i U_{i}^{model} \hat{\nu}_i
\end{align}
containing only one-body and two-body operators in the $\mathcal{D}$ subspace; $\hat{N}^d=\sum_{\s,d\k\in\mathcal{D}} c_{d\k\s}^{\dag}c_{d\k\s}$.  
It is the {\it unique} Hamiltonian of this form such that the minimization of 
\begin{align}
\Omega^{model}[\rho^d] = \mathrm{Tr}[(\hat{H}^{model}-\mu \hat{N}^d-\tau\hat{S}^d)\hat{\rho}^d] {,} \label{eq:Omega:model}
\end{align}
on the ER domain yields the exact equilibrium $\rho_1^d$ and $\rho_2^d$.  The existence of $\hat{H}^{model}$ is presently a working assumption \footnote{We recently became aware of work \cite{gonze2018} extending the exact factorization concept \cite{hunter1975,gidopoulos2014,abedi2010} to a Fock space representation and remark that at $\tau=0$ one could replace the dependence on $\rho_2^d$ in our energy functional by a dependence on the marginal function $p_J$ defined therein, where $J=(j_1\s_1,j_2\s_2,\ldots)$ can here be taken to label all possible 1-body, 2-body, $\ldots$, $N$-body states $|J>=c_{j_1\s_1}^{\dag} \ldots c_{j_N\s_N}^{\dag} |0>$ that can be constructed from orbitals in the $\mathcal{D}$ subspace.  In this case, there exists a model Hamiltonian whose ground state is $\sum_{J} p_{J} |J>$.} but appears to be true for the two-orbital Hubbard model studied in the following section.

The model Hamiltonian is a functional of the density, paramagnetic current density and the strongly correlated natural orbitals $\phi_{d\k}$.  Equations~(\ref{eq:GKS}), (\ref{eq:n}) and the minimization of $\Omega^{model}$ in Eq.~(\ref{eq:Omega:model}) lead to a set of coupled equations whose self-consistent solution returns the equilibrium $(n,\mathbf{j}_p,\phi_{d\k},f_{d\k},\rho_2^d)$.  The model Hamiltonian can be identified with a lattice model by Fourier transforming from momentum space to real space.  Assuming Born-von K\'arm\'an boundary conditions, the number of lattice sites is equal to the number of primitive cells times the number of bands.  One can perform numerical calculations for finite $\k$ point grids (a finite number of primitive cells) and extrapolate to the thermodynamic limit to obtain the result for the infinite crystal.  In the following sections, we demonstrate the viability of the above theory by constructing $\hat{H}^{model}$ and evaluating how it depends on the density in a two-orbital Hubbard model.

\section{\label{sec:application} Two-orbital Hubbard model}

The effectiveness of the theory will depend on the ability to find accurate density functionals for $v_{hxc}(\r)$ and $w_{d\k,d'\k'}$ in Eq.~(\ref{eq:h}) as well as the parameters in $\hat{H}^{model}$.  Assuming that $v_s(\r)$ can be approximated by an existing semilocal DFT approximation and that $w_{d\k,d'\k'}$ has a relatively weak density dependence, the key issue is finding functional approximations for $\hat{H}^{model}$.  If the density dependence of the model parameters is too strong or pathological, $\hat{H}^{model}$ will be difficult to approximate.  To investigate this issue, we perform numerically exact calculations for a one-dimensional two-orbital Hubbard model.  The model is defined on a bipartite lattice with each atom hosting two atomic orbitals, an $s$ orbital and a $d$ orbital.  Since the $s$ orbitals are assumed to be noninteracting and the $d$ orbitals feel a strong on-site Hubbard interaction, the model forms two strongly correlated and two weakly correlated natural occupation number bands.  The difference $n_s = (n_{Bs}-n_{As})/2$ in the $s$-orbital occupation on the $A$ and $B$ sublattices serves as a representative for the density $n(\r)$ in the continuum case, and we investigate how strongly the effective model Hamiltonian for the $d$ bands depends on $n_s$ at temperature $\tau=0$.

The Hamiltonian of our two-orbital Hubbard model is 
\begin{align}
\hat{H} &= \hat{H}_s + \hat{H}_d + \hat{H}_{sd} \label{eq:H}
\end{align}
with
\begin{align}
\hat{H}_s &= -\sum_{i\s} (t_{s,ii+1}(\xi) c_{i\s}^{\dag} c_{i+1\s} + H.c.) + \sum_{i\s} \epsilon_{s,i} c_{i\s}^{\dag} c_{i\s} \nn \\
\hat{H}_d &= -\sum_{i\s} (t_{d,ii+1}(\xi) d_{i\s}^{\dag} d_{i+1\s} + H.c.) + \sum_{i\s} \epsilon_{d,i} d_{i\s}^{\dag} d_{i\s}  \nn \\
&\quad+ U \sum_i \hat{n}_{d,i\u} \hat{n}_{d,i\d} \nn \\   
\hat{H}_{sd} &= -t_{sd} \sum_{i\s} (c_{i\s}^{\dag} d_{i+1\s} + H.c. + d_{i\s}^{\dag} c_{i+1\s} + H.c.) {,} \label{eq:H:terms}
\end{align}
where $c_{i\s}^{\dag}$ and $c_{i\s}$ are the creation and annihilation operators for an electron in the $s$ orbital at site $i$ and $d_{i\s}^{\dag}$ and $d_{i\s}$ are the corresponding operators for the $d$ orbital.  Odd sites correspond to $A$ atoms and even sites to $B$ atoms.  The staggered on-site potentials are 
\begin{align}
\epsilon_{s,i} &=(-1)^{i-1} \Delta_s \nn \\
\epsilon_{d,i} &= (-1)^{i-1} \Delta_d {,}
\end{align}
and the hopping amplitudes of the dimerized bonds are 
\begin{align}
t_{s,ii+1}(\xi) &= \left\{ \bar{ll} t_{s1} = t_{s0}- 2g_s \xi  &\quad i=\textrm{odd} \\ 
t_{s2} = t_{s0} + 2g_s \xi  &\quad i=\textrm{even} \ear \right. {,} 
\end{align}
and similarly for $t_{d,ii+1}$.  Here, $\xi$ denotes the displacement of sublattice $B$ with respect to sublattice $A$ and $g_{\alpha}$ is the electron-phonon coupling of orbital $\alpha$.  Unlike in other models of strongly correlated electrons, the $d$ bands are not assumed to be narrower than the $s$ bands.  In fact, we set $t_{s0}=t_{d0}$, so that the paramagnetic mean-field energy bands shown in Fig.~\ref{fig:bands:energy} cannot be separated into narrow low-energy bands near the Fermi energy and high-energy bands farther away, as they overlap energetically in a large region of the Brillouin zone.  $\hat{H}_{sd}$ describes a nearest-neighbor $s$-$d$ hybridization; on-site $s$-$d$ hybridization is forbidden by symmetry.  We shall show that the natural occupation number bands can be unequivocally separated into strongly- and weakly correlated sets even when the mean-field energy bands, representative of KS energy bands, do not separate into sets of wide and narrow bands.  In contrast to a common view, a partitioning into strongly  and weakly correlated bands does not require some mean-field bands to be narrower than others.

All our calculations are performed at half-filling for a supercell consisting of three primitive cells (supercell length $L=3a$ with lattice constant $a$), so there are 12 electrons occupying 12 orbitals (6 sites $\times$ 2 orbitals/site). The many-body basis and Hamiltonian were generated with the SNEG program \cite{zitko2011}.

\section{\label{sec:bandstructure} Natural occupation number band structure}

\begin{figure}[t!]
\centering
\includegraphics[width=0.85\columnwidth]{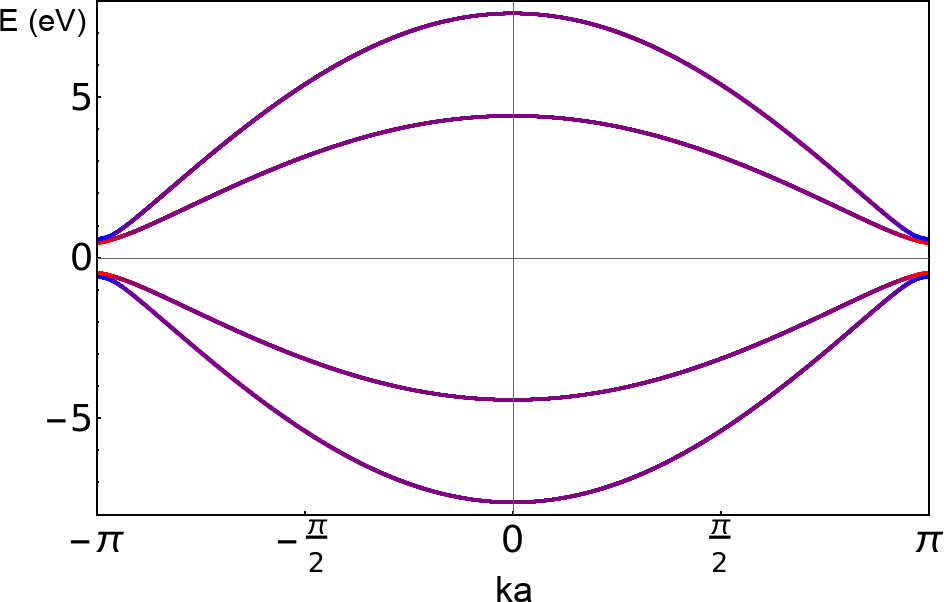}
\caption{Mean-field energy bands of the two-orbital Hubbard model versus $ka$ for $t_{s0}=t_{d0}=3$, $g_s=g_d=10$, $\Delta_s=\Delta_d=0.5$, $\xi a=0.0080$, $t_{sd}=0.8$, and $U=2.0$ (all in eV).  The strongly hybridized bands are colored according to their orbital character (blue = $s$, red = $d$, purple = hybridized).} 
\label{fig:bands:energy}
\end{figure}

\begin{figure}[h!]
\centering
\includegraphics[width=0.99\columnwidth]{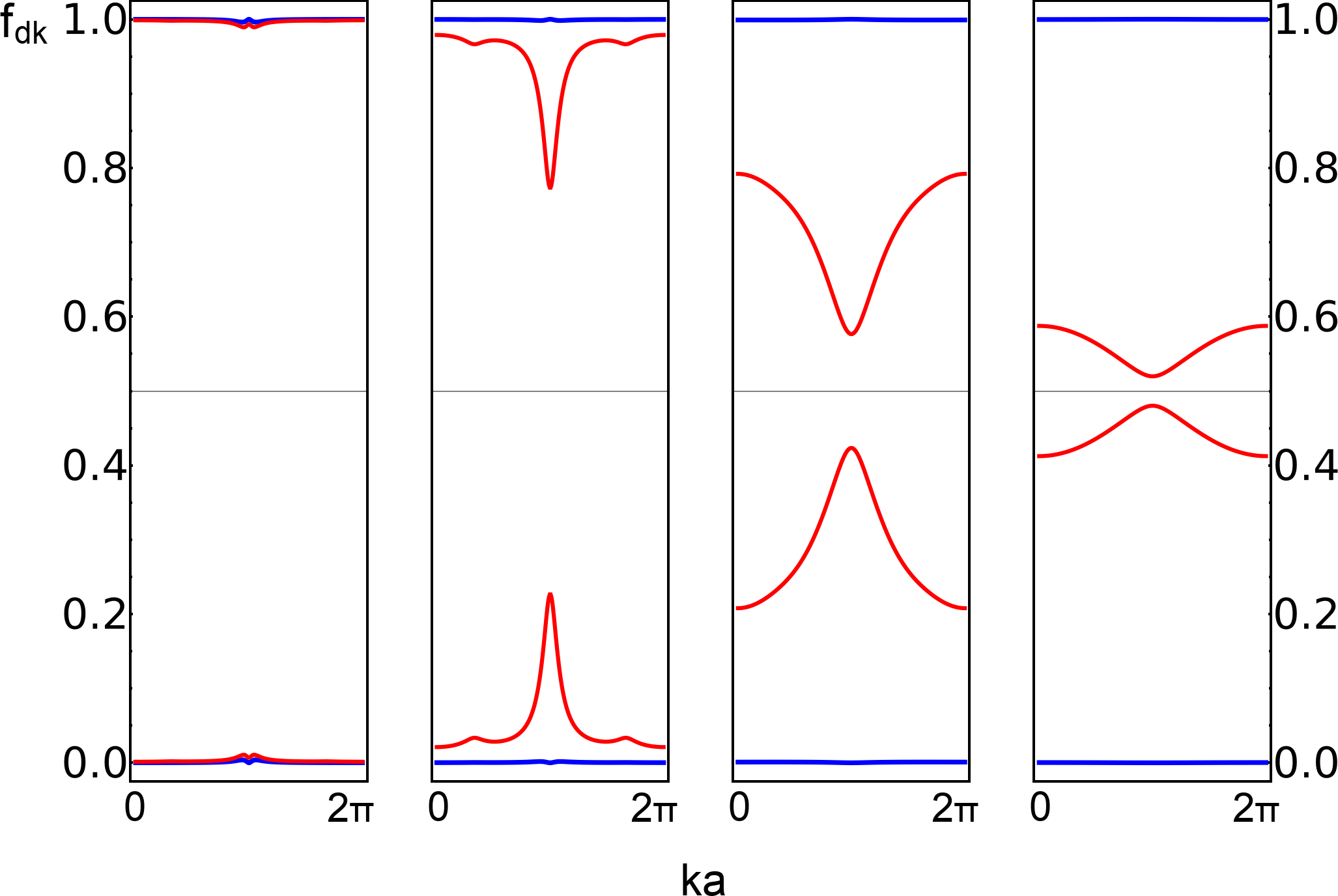} 
\caption{Natural occupation number bands of the two-orbital Hubbard model are shown for the series $U/t_{d0}=0.5,2,8,32$ (left to right) for $t_{s0}=t_{d0}=3$, $g_s=g_d=10$, $\Delta_s=\Delta_d=0.5$, $\xi a=0.0080$ and $t_{sd}=0.8$ (all in eV).  The predominantly $s$-orbital bands (blue) are close to 0 or 1 for all $U$, while the predominantly $d$-orbital bands (red) split off as $U$ increases.  The bands are shown on the domain $ka\in(0,2\pi)$ so that zone boundary is positioned at the center.} 
\label{fig:bands:series:U}
\end{figure}

The natural occupation number band structure is an alternative, more intrinsic single-particle picture of the crystal-symmetric electronic structure of a material that is useful in identifying strongly correlated degrees of freedom and calculating correlated geometric and topological properties.  To illustrate this point, we consider again the case shown in Fig.~\ref{fig:bands:energy}, where the mean-field Bloch states are strongly hybridized, i.e.~they are nearly equal mixtures of $s$ and $d$ orbitals, and relatively unaffected by $U$, provided we insist on maintaining the spin-symmetry of the problem.  The natural occupation number bands in Fig.~\ref{fig:bands:series:U} display a strikingly different behavior.  Already for moderate interaction strength $U/t_{d0}=2$, there is a clear separation into two predominantly $d$-character bands (red curves) with occupation numbers significantly different from 0 and 1 and two predominantly $s$-character bands (blue curves) with occupation numbers near 0 and 1.  The $d$ bands split off further as $U$ increases, becoming more strongly correlated as the occupation numbers approach $0.5$;  the deviation from $0$ and $1$ is a measure of the strength of correlation.  Our fundamental assumption is that the density of the weakly correlated (blue) bands is already well-described by semilocal DFT functionals, while the strongly correlated bands are better described at a higher level of theory.  
Indeed, since the occupation numbers of the weakly correlated bands are very close to 0 and 1, the many-body wavefunction approximately factors into the product of a Slater determinant of weakly correlated orbitals and a strongly correlated wavefunction in the $\mathcal{D}$ subspace, suggesting that a semilocal DFT approximation will provide a sufficiently accurate approximation for the weakly correlated bands.  Smooth and continuous natural occupation number bands and natural Bloch orbitals in the true Brillouin zone of the crystal are defined, as described in Ref.~\onlinecite{requist2018}, by unfolding the bands obtained from the full many-body wavefunction under twisted boundary conditions \cite{thouless1983,niu1984} or, equivalently, artificial magnetic fields \cite{kohn1964}.   

\begin{figure}[t!]
\centering
\includegraphics[width=0.88\columnwidth]{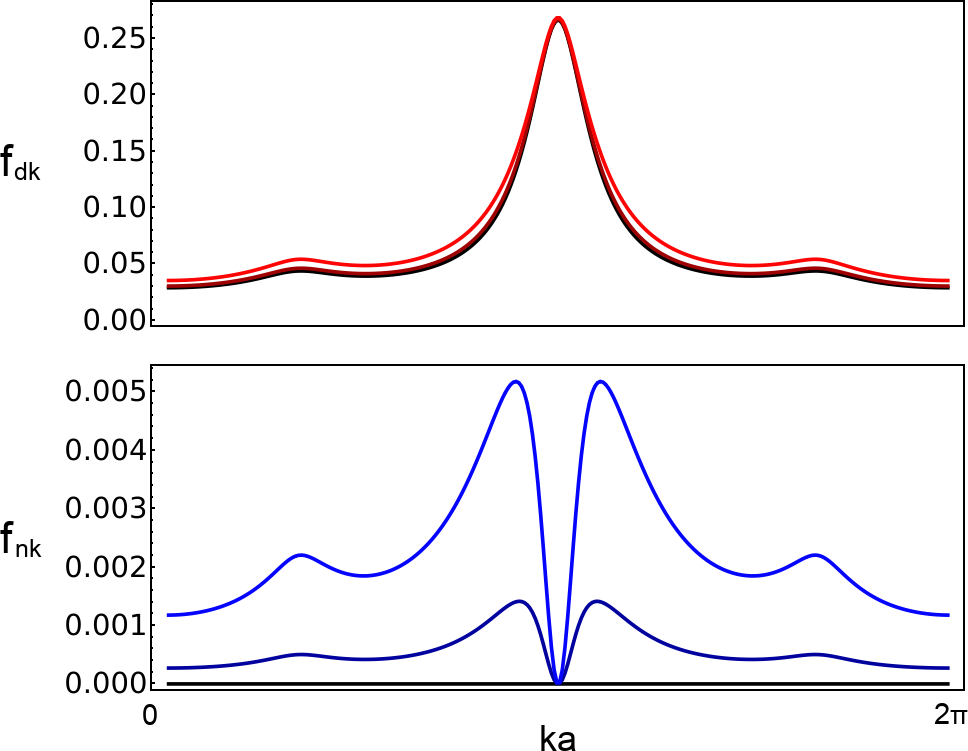} 
\caption{Lower natural occupation number bands of the two-orbital Hubbard model are shown for the series $t_{sd}=0,0.8,1.6$ (dark to light) for $t_{s0}=t_{d0}=3$, $g_s=g_d=10$, $\Delta_s=\Delta_d=0.5$, $\xi a=0.0080$ and $U/t_{d0}=2.4$; all parameters in eV.} 
\label{fig:bands:series:tsd}
\end{figure}

If we express the Hubbard interactions in Eq.~(\ref{eq:H:terms}) in the basis of the $s$-$d$ hybridized eigenstates of the noninteracting part of $\hat{H}$, we generate Hubbard interactions between $s$-type natural Bloch orbitals as well as interorbital interactions between $s$-type and $d$-type natural Bloch orbitals.  Therefore, the $s$-type natural occupation number bands begin to feel the interactions and begin to correlate when the hybridization $t_{sd}$ is turned on.  Figure \ref{fig:bands:series:tsd} shows that the natural occupation numbers of the $s$ bands correspondingly deviate from 0 and 1 and that this effect increases with increasing $t_{sd}$.  However, these deviations still remain much smaller than those of the $d$ bands even for $t_{sd}$ as large as 1.6, and hence we still have a clear separation between weakly  and strongly correlated occupation number bands.

\section{\label{sec:model} Effective model Hamiltonian}

A model Hamiltonian whose minimum yields the occupation numbers $f_{dk}$ and two-body reduced density matrix $\rho_2^d$ in the strongly correlated subspace was introduced in Sec.~\ref{sec:theory}.  $f_{dk}$ and $\rho_2^d$ generally contain contributions from many-body states with all possible occupations of the $d$ bands.  As a consequence of the particle-hole symmetry of the two-orbital Hubbard model, the natural occupation number bands have reflection symmetry about $1/2$, i.e.~for each $f_{dk}\leq 1/2$, there is another related band with occupation number $1-f_{dk}$ \cite{requist2018}.  Hence, for our chosen supercell, the mean number of electrons occupying the strongly correlated subspace is an integer, $N_d=6$.  Moreover, in the large $U$ regime, interband fluctuations are suppressed, so that states with $N_d\neq 6$ occur with low probability.  Therefore, we will attempt to find the model Hamiltonian $\hat{H}^{model,N_d}$ with a $6$-electron ground state $|\Phi_0^{N_d}\rangle$ that contracts to $f_{dk}$ and $\rho_2^d$.

To make the numerical calculation of the full $|\Psi_0\rangle$ manageable, we solve the problem in a restricted but relevant Hilbert space spanned by the sectors with $N_d=5$, 6, and 7.  More precisely, we include the following three types of states: (a) $N_d=6$ states $|\Psi_a\rangle = |S_0\rangle \otimes |D_{600}\rangle$, (b) $N_d=5$ states $|\Psi_b\rangle = a_{u,-k,-\s}^{\dag} |S_0\rangle \otimes |D_{5k\s}\rangle$ and (c) $N_d=7$ states $|\Psi_c\rangle = a_{v,k,\s} |S_0\rangle \otimes |D_{7k\s}\rangle$, where $|D_{N_dK_d S_{zd}}\rangle$ denotes an $N_d$-electron state with total quasimomentum and spin quantum numbers $K_d$ and $S_{zd}$, $|S_0\rangle = \prod_{vk\s} a_{vk\s}^{\dag} |0\rangle$ is the six-electron Fermi sea of $s$ electrons; $a_{v,k,\s}$ and $a_{u,k,\s}^{\dag}$ are the annihilation and creation operators for valence ($v$) and conduction ($u$) band $s$-electron Bloch states.  In other words, we only allow a single particle- or hole-type excitation with respect to the reference states of type (a).  This is a good approximation in all of our calculations because $U$ is large and $t_{sd}$ is relatively small.

As another consequence of particle-hole symmetry, the only two independent density variables are $n_s = \f{1}{2} (n_{Bs}-n_{As})$ and $n_d = \f{1}{2}(n_{Bd}-n_{Ad})$.  In particular, this means that if we write $\hat{H}^{\mathit{model}}$ in the natural orbital basis as
\begin{align}
\hat{H}^{model} &= \sum_{12} \epsilon_{12} c_{d_1 k_1 \s_1}^{\dag} c_{d_2 k_2 \s_2} \nn \\
&+ \sum_{1234} U_{1234} c_{d_1 k_1 \s_1}^{\dag} c_{d_2 k_2 \s_2}^{\dag} c_{d_4 k_4 \s_4} c_{d_3 k_3 \s_3} {,} \label{eq:Hmodel:Ansatz}
\end{align}
then all of the Hamiltonian parameters are functions of $n_s$ and $n_d$.  Here, $c_{dk\s}^{\dag}$ is the creation operator for an electron in the natural Bloch orbital state $\phi_{dk\s}$ and $1=(d_1 k_1 \s_1)$.  A quite general approximate form for the two-body model parameters is 
\begin{align}
U_{1234} &= \iint \phi_1^*(\r) \phi_2^*(\r') \f{g_{1234}([n],\r,\r')}{\epsilon_0|\r-\r'|} \phi_3(\r) \phi_4(\r') d\r d\r' {,}
\end{align}
where ``screening'' is described by the density-dependent factor $g_{1234}([n],\r,\r')$ and by the self-consistent optimization of the orbitals $\phi_{d\k\s}(\r)$.

Alternatively, we can also express Eq.~(\ref{eq:Hmodel:Ansatz}) in the basis of unique natural Wannier functions \cite{requist2018} as
\begin{align}
\hat{H}^{\mathit{lattice}} &= -\sum_{ij,ab} \sum_{\s} t_{ij}^{ab} c_{ia\s}^{\dag} c_{jb\s}  \nn \\
&\quad+ \sum_{ijkl,abcd} \sum_{\s\tau} U_{ijkl}^{abcd,\s\tau} c_{ia\s}^{\dag} c_{jb\tau}^{\dag} c_{ld\tau} c_{kc\s} +
\cdots  \label{eq:Hlattice}
\end{align}
where $c_{ia\s}^{\dag}$ creates an electron in the natural Wannier state $w_{ia}(\r)$ of orbital $a$ at site $i$.

\subsection{\label{sec:practical} Practical downfolding scheme} 

After finding the ground state $|\Psi_0\rangle$ of the two-orbital Hubbard model and the density matrix $\hat{\rho}=|\Psi_0\rangle\langle \Psi_0|$, we use the L\"owdin partitioning technique \cite{loewdin1962} as an efficient means of inferring the model Hamiltonian.  Define $\hat{P}$ to be the projector on $12$-electron states that consist of a fully occupied valence $s$ band (six electrons) and all possible $N_d=6$ states.  The number of such states with total quasimomentum $K=0$ and spin quantum number $S_z=0$ is 136.  Define $\hat{Q}$ to be the projector onto the orthogonal complement, comprising the sectors with $N_d=5$ and $N_d=7$.  The space spanned by $\hat{Q}$ has a total of $1200$ states with $K=0$ and $S_z=0$.  The effective Hamiltonian in the $\hat{P}$ sector is
\begin{align}
\hat{H}^{\mathit{eff}} = \hat{P} \hat{H} \hat{P} + \hat{P} \hat{H} \hat{Q} (E-\hat{Q}\hat{H}\hat{Q})^{-1} \hat{Q} \hat{H} \hat{P} {.} \label{eq:Heff}
\end{align}
This Hamiltonian is still exact.  Since it depends nonlinearly on $E$, it is not limited to the ground state and has a self-consistent solution for each eigenvalue $E_n$.  For the purpose of obtaining the most important frequency-independent model parameters of a Hamiltonian $\hat{H}^{model,N_d}$ whose ground state is $|\Phi_0^{N_d}\rangle$, one can substitute $E=E_0$ in $\hat{H}^{\mathit{eff}}$ and perform a fitting as described in the next section.

\subsection{Density dependence of the model parameters}

\begin{figure}[t!]
\centering
\includegraphics[width=0.96\columnwidth]{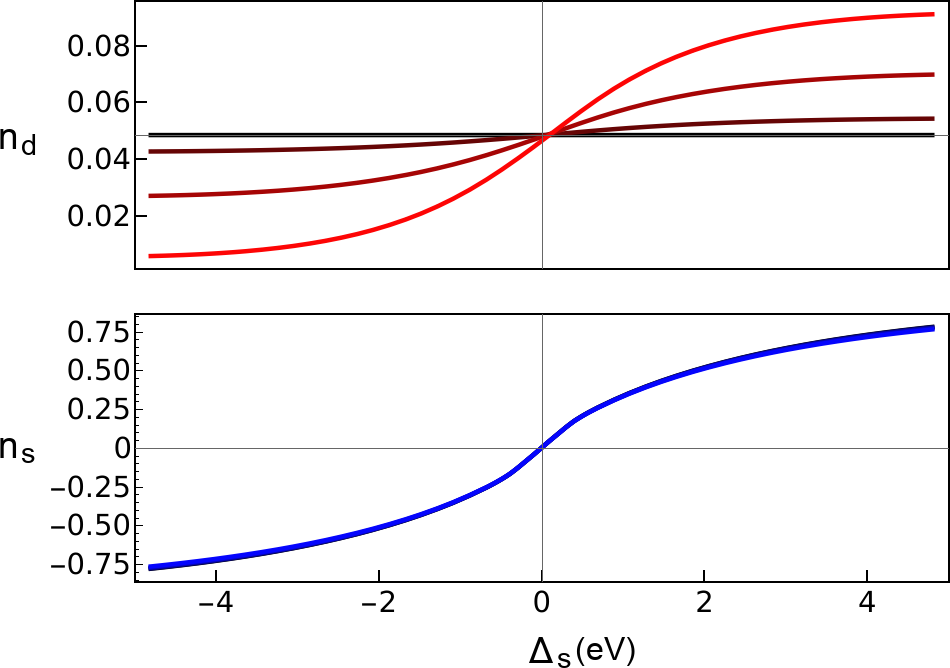}
\caption{The densities $n_d$ (red) and $n_s$ (blue) are plotted versus $\Delta_s$ for $t_{s0}=t_{d0}=3$, $g_s=g_d=10$, $\xi a=0.0020$, $\Delta_d=0.5$, $U/t_{d0}=2.4$ and for the series $t_{sd}=0.0,0.8,1.6,2.4$ (dark to light); all parameters in eV.} 
\label{fig:nd-ns-vs-DeltaS}
\end{figure}

\begin{figure}[h!]
\centering
\includegraphics[width=0.99\columnwidth]{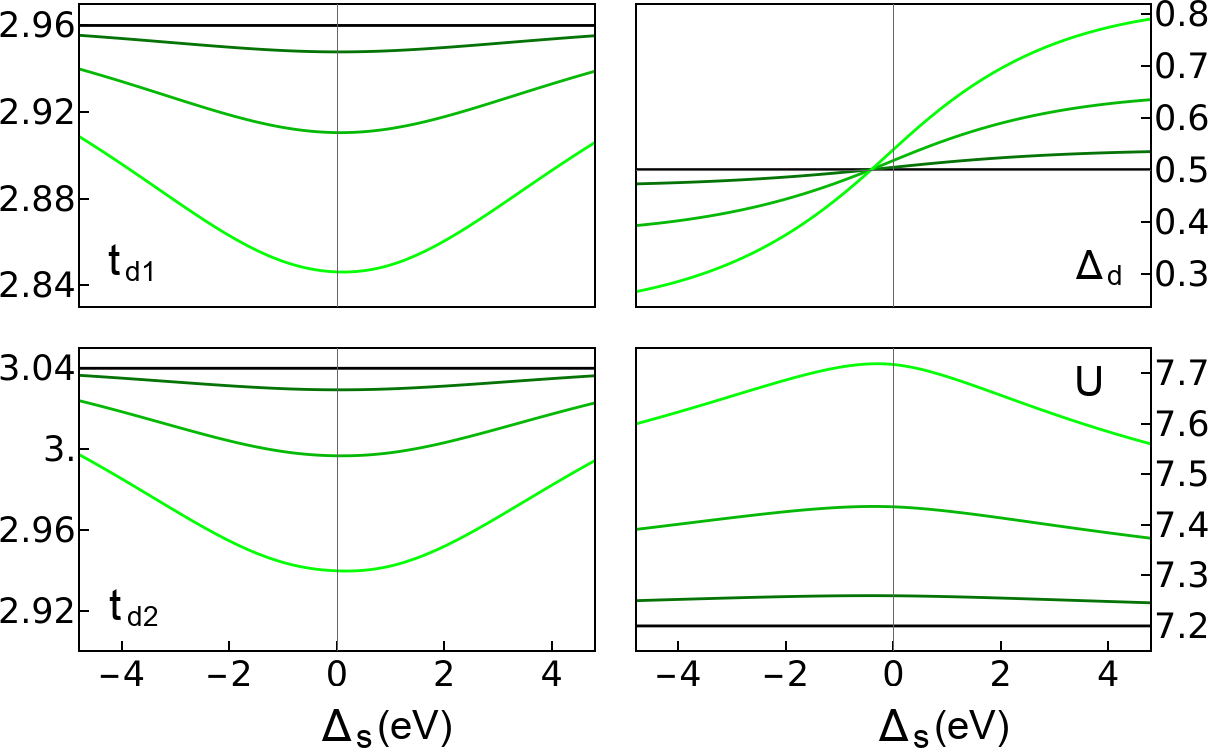}
\caption{Model parameters $t_{d1}^{model}$ (top left), $t_{d2}^{model}$ (bottom left), $\Delta_{d}^{model}$ (top right) and $U^{model}$ (bottom right) are plotted versus $\Delta_s$ for the series $t_{sd}=0.0,0.8,1.6,2.4$ (dark to light) for the same parameters as Fig.~\ref{fig:nd-ns-vs-DeltaS}.} 
\label{fig:params-vs-DeltaS}
\end{figure}

The model Hamiltonian obtained from Eq.~(\ref{eq:Heff}) is an {\it a priori} unstructured 136$\times$136 matrix on the $N_d$-electron Hilbert space with $K=0$ and $S_z=0$.  For practical calculations, we need to identify a few relevant model parameters to approximate by density functionals.  To do so, we first write a trial Hamiltonian in the original site basis
\begin{align}
\hat{H}^{model} &= -t_{d1}^{model} \sum_{i=1}^3 \sum_{\s} (d_{2i-1\s}^{\dag} d_{2i\s} + H.c.) \nn \\
&\quad- t_{d2}^{model} \sum_{i=1}^3 \sum_{\s} (d_{2i\s}^{\dag} d_{2i+1\s} + H.c.) \nn \\
&\quad+ \Delta_{d}^{model} \sum_{i=1}^3 \sum_{\s} (\hat{n}_{2i-1\s} - \hat{n}_{2i\s}) \nn \\
&\quad+ U^{model} \sum_{i=1}^6 \hat{n}_{i\u} \hat{n}_{i\d} {.}
\label{eq:Htrial}
\end{align}
To determine the model parameters, we perform a least squares minimization of $||\hat{H}^{model}-\hat{H}^{\mathit{eff}}||$ in the Frobenius norm.  Then we verify that the ground state of $\hat{H}^{model}$ is close to the ground state of $\hat{H}^{\mathit{eff}}$.  In all the calculations we report, $1-|\langle\Phi^{model,N_d}_{0} | \Phi^{\mathit{eff},N_d}_{0}\rangle|\lesssim 5\times 10^{-4}$.  Varying $\Delta_s$ and $\Delta_d$ in the original Hamiltonian allows us to change $n_s$ and $n_d$ over a range of values and study the resulting trends in the model parameters.  Figure~\ref{fig:nd-ns-vs-DeltaS} shows the one-to-one relationship between $\Delta_s$ and $n_s$.  Figure~\ref{fig:params-vs-DeltaS} shows how the model parameters vary as functions of $\Delta_s$.  By inverting the $\Delta_s\rightarrow n_s$ mapping in Fig.~\ref{fig:nd-ns-vs-DeltaS}, we can infer the density dependence of the model parameters.  The $n_s$-dependence of $t_{d1}^{model}$ and $t_{d2}^{model}$ is negligible for $t_{sd}\lesssim 1$~eV and relatively weak in all cases considered. 
\begin{figure}[h!]
\centering
\includegraphics[width=0.96\columnwidth]{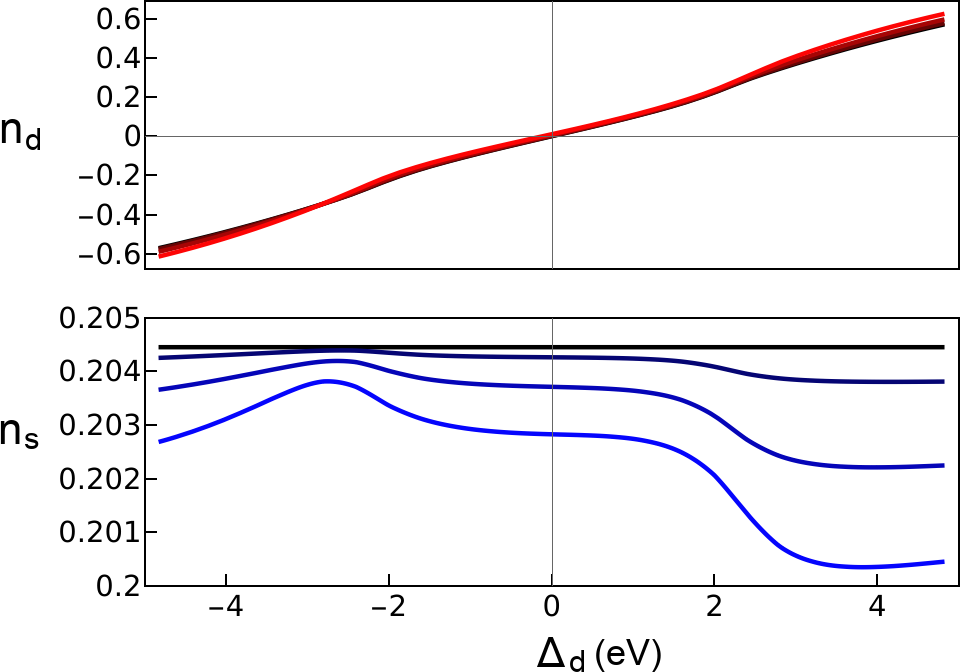}
\caption{The densities $n_d$ (red) and $n_s$ (blue) are plotted versus $\Delta_d$ for $t_{s0}=t_{d0}=3$, $g_s=g_d=10$, $\xi a=0.0020$, $\Delta_s=0.5$, $U/t_{d0}=2.4$ and for the series $t_{sd}=0.0,0.8,1.6,2.4$ (dark to light); all parameters in eV.} 
\label{fig:nd-ns-vs-DeltaD}
\end{figure}
\begin{figure}[h!]
\centering
\includegraphics[width=0.99\columnwidth]{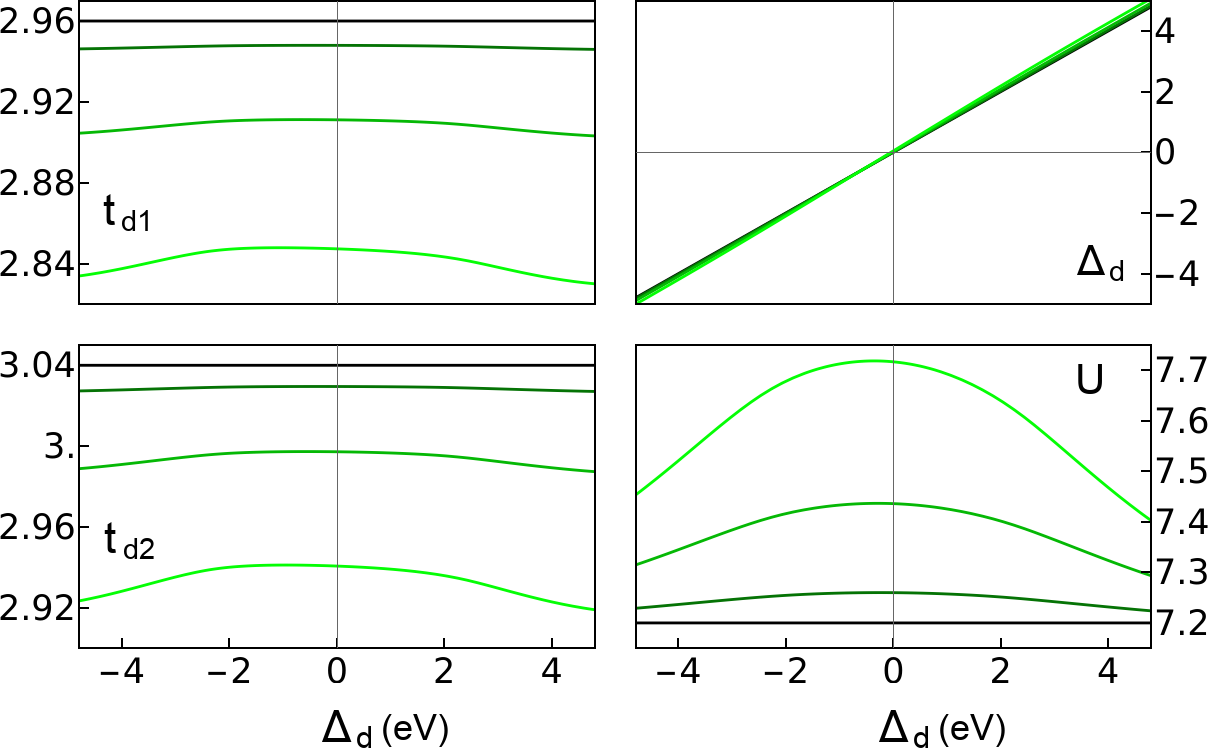}
\caption{Model parameters $t_{d1}^{model}$ (top left), $t_{d2}^{model}$ (bottom left), $\Delta_{d}^{model}$ (top right) and $U^{model}$ (bottom right) are plotted versus $\Delta_d$ for the series $t_{sd}=0.0,0.8,1.6,2.4$ (dark to light) and the same parameters as Fig.~\ref{fig:nd-ns-vs-DeltaD}.} 
\label{fig:params-vs-DeltaD}
\end{figure}
We emphasize that $t_{d1}^{model}$ and $t_{d2}^{model}$ change very little even though $\Delta_s$ spans a very large range in Figs.~\ref{fig:nd-ns-vs-DeltaS} and \ref{fig:params-vs-DeltaS}.  $\Delta_{d}^{model}$ and $U^{model}$ have moderate but still quite regular $n_s$-dependence.  Figures \ref{fig:nd-ns-vs-DeltaD} and \ref{fig:params-vs-DeltaD} show the dependence of $n_s$, $n_d$ and the model parameters on $\Delta_d$.  Here, the trends in $t_{d1}^{model}$, $t_{d2}^{model}$ and $U^{model}$ are similar to those obtained from varying $\Delta_s$.  On the other hand, $\Delta_{d}^{model}$ has an approximately linear relationship with $\Delta_d$, which is expected.  These results provide encouraging evidence that simple functional approximations can be found for a model Hamiltonian that describes a few select strongly correlated bands in real materials.  

We have considered more general trial Hamiltonians with next-nearest neighbor hopping amplitudes $t_3$ and $t_4$ and next-next-nearest neighbor hopping $t_5$, as well as several different types of two-body interactions.  As all of these additional terms are generically nonzero, we observe that the coupling of the $d$ bands to the $s$ bands generates interactions beyond the Hubbard interactions in the original Hamiltonian.  However, throughout the range of parameters reported here, these additional terms are small and the Hamiltonian in Eq.~(\ref{eq:Htrial}) is sufficient.

\section{\label{sec:polarization} Macroscopic polarization}

Having obtained a model Hamiltonian whose ground state gives a good approximation to the state of the strongly correlated subspace, we now test how well such a model Hamiltonian and ground state can reproduce the strongly correlated part of the macroscopic polarization.   The Ortiz-Martin formula $P = -(e/2\pi)  \lim_{N\rightarrow\infty}\gamma(N)$ relates the macroscopic polarization (modulo the polarization quantum) to the many-body geometric phase 
\begin{align}
\gamma(N) = \int_0^{2\pi} i \langle \Psi_0 | \partial_{\alpha} \Psi_0 \rangle d\alpha {,} \label{eq:gamma:OM}
\end{align}
where in our case $|\Psi_0(\alpha)\rangle$ is the $12$-electron ground state of the twisted version of the Hamiltonian in Eq.~(\ref{eq:H}) \cite{ortiz1994}.  The twisted Hamiltonian is obtained from $\hat{H}$ by making the Peierls's substitutions $t_{aj}\rightarrow t_{aj} e^{i\alpha/6}$ (with $a=s,d$ and $j=1,2$) and $t_{sd}\rightarrow t_{sd} e^{i\alpha/6}$.

To evaluate Eq.~(\ref{eq:gamma:OM}), we express the wavefunction as
\begin{align}
|\Psi_0\rangle = \sum_D c_D |S(D)\rangle |D\rangle {,}
\label{eq:Psi:restricted}
\end{align}
where $D=(d_1k_1\s_1,d_2k_2\s_2,\ldots)$ is a multi-index labeling the strongly correlated $d$-orbital part of the many-body state.  The wavefunction in Eq.~(\ref{eq:Psi:restricted}) has a restricted form because we have solved the problem on the restricted Hilbert space consisting of the $N_d=$ 5, 6, and 7 sectors defined in Sec.~\ref{sec:model}.  Hence, the $s$-electron factor, denoted as $|S(D)\rangle$, is uniquely determined by $|D\rangle$.  For example, for a state $|D\rangle$ with $N_d=7$, $K_d=k$ and $S_{zd}=\sigma$, conservation of particle number, quasimomentum and $z$-projection of spin imply $|S(D)\rangle = a_{vk\sigma} |S_0\rangle$.  Thus, for our system, the geometric phase in Eq.~(\ref{eq:gamma:OM}) is
\begin{align}
\gamma &= \sum_D \int_0^{2\pi} i c_D^* \partial_{\alpha} c_D d\alpha \nn \\
&+ \sum_{N_dK_dS_{zd}} \int_0^{2\pi} i P_{N_dK_dS_{zd}} \sum_{nk\s} f_{nk\s}^{N_dK_dS_{zd}} \langle \phi_{nk\s} | \partial_{\alpha} \phi_{nk\s} \rangle d\alpha {,} \label{eq:gamma:1}
\end{align}
where $P_{N_dK_dS_{zd}}(\alpha) = \sum_{D}^{(N_dK_dS_{zd})} |c_D(\alpha)|^2$ and $f_{nk\s}^{N_dK_dS_{zd}}$ is the occupation number (0 or 1) of the $s$-orbital $\phi_{nk\s}$ in the Slater determinant $|S(D_{N_dK_dS_{zd}})\rangle$.  After unfolding the natural occupation numbers and natural Bloch orbitals to the full Brillouin zone \cite{requist2018}, the geometric phase simplifies to 
\begin{align}
\gamma &= \int_0^{2\pi} i \langle \Phi_0^{N_d} | \partial_{\alpha} \Phi_0^{N_d} \rangle d\alpha \nn \\
&\quad+ \sum_{\s,n\in\mathcal{S}} \int_0^{2\pi/a} i f_{nk\s} \langle \tilde{u}_{nk\s} | \partial_k \tilde{u}_{nk\s} \rangle dk {.} \label{eq:gamma:2}
\end{align}
The first term has been expressed in terms of the ground state $|\Phi_0^{N_d}\rangle$ of the twisted model Hamiltonian, while the second term is given in terms of the periodic part $\tilde{u}_{nk\s}$ of the $s$-electron Bloch orbital $\phi_{nk\s}$.  Given that the occupation numbers $f_{nk\s}=\sum_{N_dK_dS_{zd}} P_{N_dK_dS_{zd}} f_{nk\s}^{N_dK_dS_{zd}}$ of the $s$-electron Bloch orbitals are close to $0$ or $1$, we suppose that the second term can be well-approximated by the geometric phase of noninteracting electrons as in the King-Smith--Vanderbilt formula, i.e. 
\begin{align}
\gamma_{s}^{KS} = \sum_{n\s} \int_0^{2\pi/a} i g_{nk\s} \langle u_{nk\s} | \partial_k u_{nk\s} \rangle dk {,} \label{eq:gamma:s}
\end{align}
where the occupation numbers $g_{nk\s}$ (0 or 1) restrict the sum to the occupied $s$-electron Kohn-Sham-like states $\chi_{nk\s}(r)=e^{ikr} u_{nk\s}(r)$.  Although we expect Eq.~(\ref{eq:gamma:s}) to be an accurate approximation for the weakly correlated bands in line with the successful application of the King-Smith--Vanderbilt formula to weakly correlated systems, the polarization calculated by Eq.~(\ref{eq:gamma:2}) with the approximation in Eq.~(\ref{eq:gamma:s}) is not exact.  It might be possible to extend our theory to an exact theory for the macroscopic polarization by including additional basic variables in analogy to the inclusion of the polarization in standard DFT \cite{gonze1995,gonze1997,martin1997}.

\begin{figure}[t!]
\centering
\includegraphics[width=0.8\columnwidth]{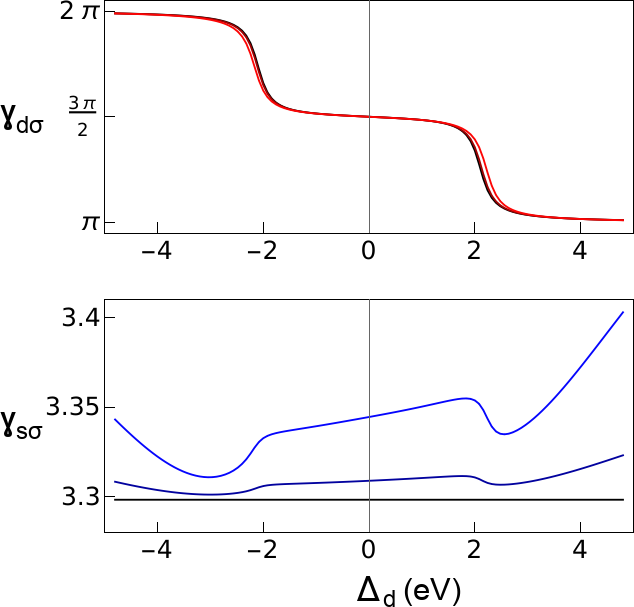} 
\caption{Many-body geometric phases $\gamma_{d\s}$ (red) and $\gamma_{s\s}$ (blue) for the  two-orbital Hubbard model are plotted versus $\Delta_d$ for the series $t_{sd}=0,0.8,1.6$~eV (dark to light) and the same parameters as Fig.~\ref{fig:nd-ns-vs-DeltaD}.} 
\label{fig:gammaSD:WNH}
\end{figure}

\begin{figure}[t!]
\centering
\includegraphics[width=0.8\columnwidth]{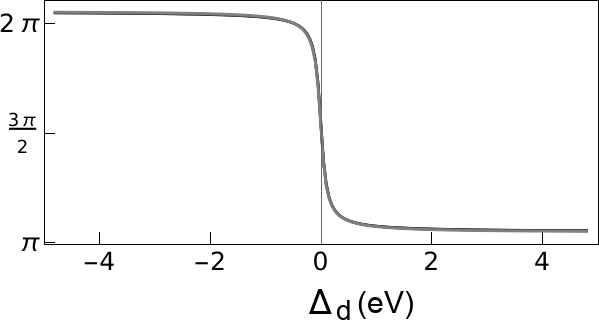} 
\caption{Geometric phase $\gamma_{d\s}+(\gamma_{s\s}-\pi)$ for the noninteracting two-orbital Hubbard model is plotted versus $\Delta_d$ for the series $t_{sd}=0,0.8,1.6$ (dark to light gray) and the same parameters as Fig.~\ref{fig:gammaSD:WNH}.  The curves for different values of $t_{sd}$ nearly coincide.} 
\label{fig:gammaSD:WN}
\end{figure}

We begin by presenting results for the exact geometric phase calculated with Eq.~(\ref{eq:gamma:2}). 
The first term of Eq.~(\ref{eq:gamma:2}), which we denote as $\gamma_d$, is the contribution of the strongly correlated $d$-electron wavefunction.  Since $\gamma_d$ has equal contributions from spin up and spin down electrons, we define $\gamma_{d\s} = \gamma_d/2$.  The second term, $\gamma_s$, is the contribution of the weakly correlated $s$-electron bands, and we further define $\gamma_{s\s}=\gamma_s/2$.  $\gamma_{d\s}$ and $\gamma_{s\s}$ are shown versus $\Delta_d$ in Fig.~\ref{fig:gammaSD:WNH}.  As the $s$ electrons only feel the bias indirectly through their hybridization with the $d$ bands, $\gamma_{s\s}$ has a weak dependence on $\Delta_d$.  On the other hand, $\gamma_{d\s}$ has a nontrivial dependence on $\Delta_d$.  The plateau between $\Delta_d =-2$ and 2~eV is due to Hubbard interactions; $\Delta_d$ can only polarize the $d$-electron states if it can overcome the on-site repulsion $U$.  The effects of correlations on $\gamma_{d\s}$ are evident upon comparison with the geometric phase of the noninteracting version ($U=0$) of the two-orbital Hubbard model, which is shown in Fig.~\ref{fig:gammaSD:WN}.

\begin{figure}[t!]
\centering
\includegraphics[width=0.85\columnwidth]{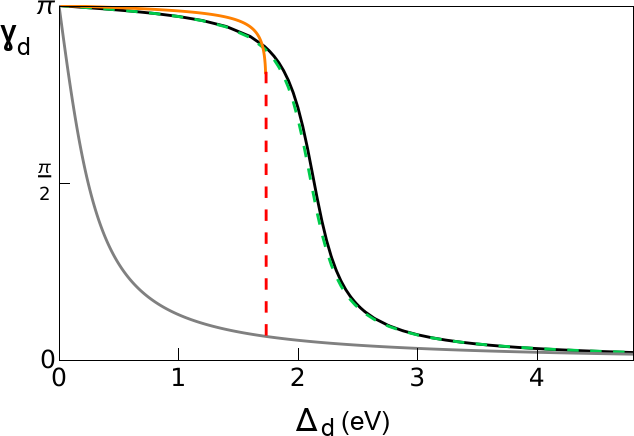} 
\caption{Comparison of $\gamma_d$ from the full solution of the two-orbital Hubbard model (black), the model Hamiltonian solution (dashed green) and mean-field approximations (orange and gray) for $t_{s0}=3$, $t_{d0}=3$, $\xi=0.0020$, $\Delta_s=0.5$, $U/t_{d0}=2.4$ and $t_{sd}=0.8$ as functions of $\Delta_d$; all parameters in eV.  The dashed red  line connects the paramagnetic (gray) and antiferromagnetic (orange) mean-field results.} 
\label{fig:geometric:phase:model-vs-meanfield}
\end{figure}

So far we have presented results for the exact $|\Psi_0(\alpha)\rangle$.  Now we ask how well the ground state $|\Phi_{0}^{model,N_d}\rangle$ of the fitted model Hamiltonian reproduces $\gamma_{d\s}$.  In principle, the model Hamiltonian parameters will be $\alpha$-dependent.  However, we can make the following approximation.  We assume that the {\it moduli} of the hopping parameters $t_{d1}^{model}$ and $t_{d2}^{model}$ are approximately constant but that their phases vary as functions of $\alpha$ exactly as the bare hopping parameters do under the Peierls's substitution.  Under this approximation, the only information we need to evaluate $\gamma_{d\s}$ are the parameters of the untwisted model Hamiltonian ($\alpha=0$). To test the validity of this approximation, we use the $\alpha=0$ model parameters from Fig.~\ref{fig:params-vs-DeltaD} to calculate $\gamma_{d,\mathit{approx}}$ (dashed green) and compare it with the exact result (black) in Fig.~\ref{fig:geometric:phase:model-vs-meanfield}.  The results are shown together with the mean-field (Hartree-Fock) approximation.  First, they demonstrate that the ground state of $\hat{H}^{model}(\alpha)$, obtained by performing the Peierls's substitution to $\hat{H}^{model}(0)$, yields an accurate approximation to the strongly correlated part of the full geometric phase.  Second, they reveal that the paramagnetic mean-field approximation (gray) fails qualitatively.  The broken-symmetry antiferromagnetic mean-field approximation (orange) improves the behavior for small $\Delta_d$ but fails for $1.75 \lesssim \Delta_d \lesssim 3.5$~eV.  Semilocal (spin-)DFT approximations will give comparably incorrect results.  

\begin{figure}[t!]
\centering
\includegraphics[width=0.85\columnwidth]{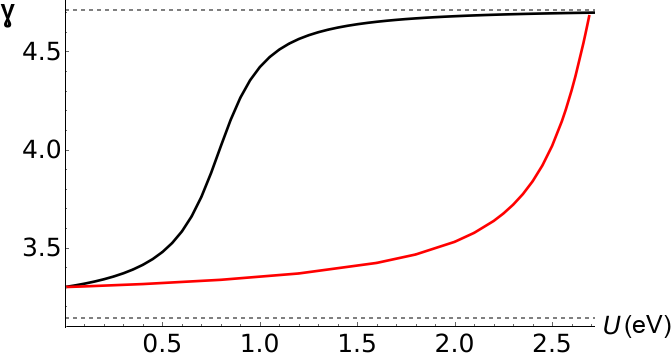} 
\caption{Comparison of BALDA (red) and exact (black) results for the geometric phase in the Rice-Mele-Hubbard as a function of $U$ for $t_{d0}=3$, $g_d=10$, $\Delta_d=0.5$ and $\xi=0.0020$; all parameters in eV.  Dashed lines show $\pi$ and $3\pi/2$.} 
\label{fig:geometric:phase:BALDA}
\end{figure}

Bethe Ansatz local density approximation (BALDA) \cite{lima2003,schoenhammer1995} uses the one-dimensional Hubbard model as a reference system from which to derive the exchange-correlation energy density as a function of the average site occupation.  The BALDA exchange-correlation energy density can be used to calculate the density of inhomogeneous lattice models such as Hubbard models with staggered potentials.  Since, by construction, the BALDA yields the exact energy and density for a uniform one-dimensional lattice model for any $U$, one might expect it to be capable of reproducing the strongly correlated part of the geometric phase.  To test this, we have evaluated the geometric phase $\gamma_{d,\mathit{BALDA}}$ of the Rice-Mele-Hubbard model by substituting the BALDA Kohn-Sham eigenstates into the King-Smith--Vanderbilt formula.  As shown in Fig.~\ref{fig:geometric:phase:BALDA}, $\gamma_{d,\mathit{BALDA}}$ agrees with the exact geometric phase for $U=0$ but is qualitatively incorrect for moderate and large $U$.  We conclude that functional approximations that yield good densities do not necessarily yield accurate values for the geometric phase, which depends on how the $\k$-dependent phases of the Bloch functions vary across the Brillouin zone.

\section{\label{sec:outlook} Conclusions}

Self-consistently coupling density functional theory to a model Hamiltonian (``DFT+model'') was shown to yield accurate results for the macroscopic polarization in a strongly correlated system where all known DFT approximations fail qualitatively.  DFT+model is an efficient {\it ab initio} framework for calculating geometric and topological invariants in strongly correlated materials.  The application of the theory to the calculation of topological invariants is a problem for future work.  The fact that the theory reliably reproduces the macroscopic polarization --- a geometric quantity closely related to topological invariants --- over a wide range of parameters, as we have observed here, is strong evidence that it will also give accurate results for topological invariants.

The theory establishes a systematic and self-consistent {\it ab initio} procedure for constructing the unique model Hamiltonian for the subset of strongly correlated degrees of freedom.  This is the only model Hamiltonian that yields the correct equilibrium occupation numbers $f_{d\k}$ and two-body reduced density matrix $\rho_2^d$ in the strongly correlated subspace.  The two-body reduced density matrix $\rho_2^d$ contains the information about all two-body correlation functions.  The model Hamiltonian does not depend on a separation of energy scales or the existence of a set of narrow mean-field energy bands.  Identifying the strongly correlated orbitals from the natural occupation number band structure is a novel way of singling out strongly correlated degrees of freedom in solids.

As opposed to $GW$+DMFT \cite{biermann2003,sun2004,tomczak2012,ayral2012,hansmann2013,tomczak2014,biermann2014} and other DMFT-based methods, our theory does not involve any frequency dependence.  This does not imply any approximation but entails certain advantages and disadvantages.  One advantage is the greater efficiency of the many-body part of our self-consistency cycle, which involves solving for the equilibrium state of a many-body Hamiltonian rather than solving a self-consistent impurity problem with frequency-dependent interactions, e.g.~$U(\omega)$, as in some DMFT implementations \cite{werner2010}.  On the other hand, the lack of frequency dependence might make it more difficult to obtain spectral functions in our theory.  It would be interesting to explore how well the excitations of the model Hamiltonian represent the true strongly correlated excitations of the system, although we emphasize that our model Hamiltonian is only guaranteed to reproduce the equilibrium occupations and two-body correlations and not any other observables.  Another important distinction with DMFT-based approaches is the fact that our theory preserves the full crystal symmetry of the original Hamiltonian including $\k$-dependent correlations, which are neglected in conventional DMFT.  The $\k$-dependence is crucial for properly evaluating geometric and topological quantities in strongly correlated materials.  Symmetry constraints on the structure of the $\k$-dependent correlations may enable improved accuracy.  Recent extensions of DMFT aim to incorporate nonlocal correlations \cite{lechermann2017}.

Adapting the present theory to finite systems in which the strongly correlated subspace is taken to be spanned by just a few natural orbitals \cite{requist2011}, such as localized orbitals in Kondo systems or hybridized transition metal orbitals in molecules, would constitute a novel embedding theory that might allow one to obtain more accurate {\it ab initio} results for systems with strong static correlation and partially circumvent the problem of memory dependence in TDDFT.  This would be especially helpful in the modeling of coupled electron-ion dynamics within exact factorization density functional theory \cite{requist2016b,li2018}.

\appendix

\section{Generalized Kohn-Sham scheme}

The role of the on-site $d$-orbital density matrix and orbital-dependent potentials in the DFT+$U$ method \cite{anisimov1991b,liechtenstein1995,anisimov1997a,pickett1998,cococcioni2005,nakamura2006,himmetoglu2013}, as well as the uncertainties arising from the strongly correlated narrow Fe bands in semilocal DFT calculations of the pressure-induced spin state crossover in Fe$_x$Mg$_{1-x}$SiO$_3$ perovskite \cite{umemoto2008}, motivated the investigation of an effective single-particle Hamiltonian for a strongly correlated Hubbard model in a reduced density matrix approach \cite{requist2008}.  As in the DFT+$U$ method, it was assumed that the nonlocal, orbital-dependent part of the effective potential would only be applied to a subset of strongly correlated degrees of freedom.  In this Appendix, we show that the part of the nonlocal effective potential acting in the strongly correlated subspace can be rigorously combined with a multiplicative Kohn-Sham potential to define a generalized Kohn-Sham Hamiltonian.

For fixed $(f_{d\k},\rho_2^d)$, the density, paramagnetic current density, and strongly correlated natural Bloch orbitals can be obtained by self-consistently solving a generalized Kohn-Sham equation.  We start by defining the following grand potential functional for a system of noninteracting electrons in the presence of local scalar and vector potentials $v_s(\r)$ and $\mathbf{A}_s(\r)$ and a nonlocal potential $w(\r\s,\r'\s')=\sum_{d\k d'\k'} w_{d\k,d'\k'} \phi_{d\k\s}(\r) \phi_{d'\k'\s'}^*(\r)$:
\begin{align}
\Omega_s[n,\mathbf{j}_p,\phi_{d\k},f_{d\k}] &= K_s[n,\mathbf{j}_p,\phi_{d\k},f_{d\k}] + \f{e}{c} \int \mathbf{A}_s(\r)\cdot \mathbf{j}_p(\r)d\r \nn \\
&+ \int \Big[\f{e^2}{2mc^2} |\mathbf{A}_s(\r)|^2+v_s(\r)-\mu\Big] n(\r) d\r \nn \\
&+ \sum_{\s\s'}\int w(\r\s,\r'\s') \rho_1(\r'\s',\r\s) d\r d\r' {.}
\label{eq:E0}
\end{align}
The ensemble kinetic-energy-entropy functional is
\begin{align}
K_s[n,\mathbf{j}_p,\phi_{d\k},f_{d\k}] &= \min_{\rho_s \rightarrow (n,\mathbf{j}_p,\phi_{d\k},f_{d\k})} \mathrm{Tr}[(\hat{T}-\tau\hat{S}_s)\hat{\rho}_s]  \nn \\ 
&= \sum_{n\k\s}^{\mathcal{S}} g_{n\k\s} \langle \chi_{n\k\s} |\hat{T} | \chi_{n\k\s}\rangle \nn \\
&+ \sum_{d\k\s}^{\mathcal{D}} f_{d\k\s} \langle \phi_{d\k\s} |\hat{T} | \phi_{d\k\s}\rangle \nn \\
&+k_B\tau\sum_{b\k} [f_{b\k} \ln f_{b\k} \nn \\
&\quad + (1-f_{b\k}) \ln(1-f_{b\k})]
{,} \label{eq:Te}
\end{align}
where $\hat{T}=-\hbar^2\nabla^2/2m$ and the sum over $b\k$ runs over both $\mathcal{S}$ and  $\mathcal{D}$ with $f_{b\k}=f_{d\k}$ for $b\k\in\mathcal{D}$ and $f_{b\k}=g_{n\k}$ for $b\k\in\mathcal{S}$.  The weakly correlated orbitals in $\mathcal{S}$ are denoted $\chi_{n\k}$, the strongly correlated orbitals in $\mathcal{D}$ are denoted $\phi_{d\k}$, and a generic eigenstate from either $\mathcal{S}$ or $\mathcal{D}$ is denoted $\psi_{b\k}$.  In Eq.~(\ref{eq:Te}), we have allowed for the possibility that the equilibrium state, even at $\tau=0$, is an ensemble state \cite{requist2008} formed from degenerate Slater determinants $|\Phi_I\rangle$ according to 
\begin{align}
\rho_s = \sum_I w_I |\Phi_I\rangle \langle \Phi_I | {;} \quad \sum_I w_I = 1 {.}
\end{align}
The unknown ensemble weights are related to the occupation numbers according to
\begin{align}
\sum_I \Theta_{n\k,I} w_I &= g_{n\k} \quad \mathrm{if} \; n\k \in \mathcal{S} \nn \\
\sum_I \Theta_{d\k,I} w_I &= f_{d\k} \quad \mathrm{if} \; d\k \in \mathcal{D} {,}
\end{align}
where $\Theta_{b\k,I}=1$ if the orbital $\psi_{b\k}$ is an element of the Slater determinant $|\Phi_I\rangle$ and $0$ otherwise \cite{dreizler1990}.  Then, we define the functional
\begin{align}
G_s[n,\mathbf{j}_p,\phi_{d\k},f_{d\k}] &= \Omega_s[n,\mathbf{j}_p,\phi_{d\k},f_{d\k}] \nn \\
&- \sum_{b\k\neq b'\k'} \lambda_{b\k,b'\k'}^s \int \phi_{b'\k'}^*(\r) \phi_{b\k}(\r) d\r {.}
\end{align}
Taking variations with respect to $\phi_{d\k}(\r)$ and $\phi_{d\k}^*(\r)$ leads to the stationary conditions
\begin{align}
f_{d\k} \big< \phi_{d'\k'} \big| \hat{H}_{s,\mathit{local}} + \hat{W}_s \big| \phi_{d\k} \big> &= \lambda_{d'\k',d\k}^s \nn \\
f_{d'\k'} \big< \phi_{d'\k'} \big| \hat{H}_{s,\mathit{local}} + \hat{W}_s \big| \phi_{d\k} \big>  &= \lambda_{d'\k',d\k}^s  {,}
\end{align}
where $\hat{H}_{s,\mathit{local}} = -\hbar^2\nabla^2/2m + (e/c) \int \mathbf{A}_s(\r) \cdot \hat{\mathbf{j}}_p(\r) d\r + \int [(e^2/2mc^2)|\mathbf{A}_s(\r)|^2+v_s(\r)-\mu] \hat{n}(\r) d\r$ and $\hat{W}_s = \iint \hat{\psi}^{\dag}(\r) w(\r,\r') \hat{\psi}(\r')$.  By subtraction, we find
\begin{align}
(f_{d\k} - f_{d'\k'}) \big< \phi_{d'\k'} \big| \hat{T} + \hat{V}_s + \hat{W}_s \big| \phi_{d\k} \big> = 0 {.} \label{eq:stat:cond:1a}
\end{align} 
Setting $\delta G_s/\delta n(\r)=0$ and $\delta G_s/\delta \mathbf{j}_p(\r)=0$ gives 
\begin{align}
v_s(\r)-\mu + \f{e^2}{2mc^2}|\mathbf{A}_s(\r)|^2 + \left.\f{\delta K_s}{\delta n(\r)}\right|_{\phi_{d\k},f_{d\k}} &= 0 \nn \\
\f{e}{c} \mathbf{A}_s(\r) + \left.\f{\delta K_s}{\delta \mathbf{j}_p(\r)}\right|_{\phi_{d\k},f_{d\k}} &= 0 {.}
\label{eq:stat:cond:1b}
\end{align}

Next, we recall the grand potential functional in Eq.~(\ref{eq:Omega}) for a system of interacting electrons in the presence of scalar and vector potentials $v(\r)$ and $\mathbf{A}(\r)$ 
\begin{align}
\Omega[n,\mathbf{j}_p,\phi_{d\k},f_{d\k},\rho_2^d] &= K_s[n,\mathbf{j}_p,\phi_{d\k},f_{d\k}] \nn \\
&+ \f{e}{c} \int \mathbf{A}(\r)\cdot \mathbf{j}_p(\r)d\r \nn \\
&+ \int \Big[\f{e^2}{2mc^2} |\mathbf{A}(\r)|^2+v(\r)-\mu\Big] n(\r) d\r \nn \\
&+ \Omega_{hxc}[n,\mathbf{j}_p,\phi_{d\k},f_{d\k},\rho_2^d]
\end{align}
and introduce
\begin{align*}
G[n,\mathbf{j}_p,\phi_{d\k},f_{d\k},\rho_2^d] &= \Omega[n,\mathbf{j}_p,\phi_{d\k},f_{d\k},\rho_2^d] \nn \\
&- \sum_{d\k,d'\k'} \lambda_{d\k,d'\k'} \int \phi_{d'\k'}^*(\r) \phi_{d\k}(\r) d\r {.}
\end{align*}
The stationary conditions for variations of $G$ with respect to $\phi_{d\k}(\r)$ and $\phi_{d\k}^*(\r)$ are
\begin{align}
f_{d\k} \big< \phi_{d'\k'} \big| \hat{H}_{\mathit{local}} \big| \phi_{d\k} \big> &+ \left.\bigg< \phi_{d'\k'} \bigg| \f{\delta \Omega_{hxc}}{\delta\phi_{d\k}^*} \bigg>\right|_{n,\mathbf{j}_p} = \lambda_{d'\k',d\k} \nn \\
f_{d'\k'} \big< \phi_{d'\k'} \big| \hat{H}_{\mathit{local}} \big| \phi_{d\k} \big> &+ \left.\bigg< \f{\delta \Omega_{hxc}}{\delta\phi_{d'\k'}} \bigg| \phi_{d\k} \bigg>\right|_{n,\mathbf{j}_p} = \lambda_{d'\k',d\k} {,}
\end{align}
where $\hat{H}_{\mathit{local}} = -\hbar^2\nabla^2/2m + (e/c) \int [\mathbf{A}(\r)+\mathbf{A}_{xc}(\r)]\cdot\hat{\mathbf{j}}_p(\r) d\r + \int [(e^2/2mc^2)|\mathbf{A}(\r)|^2+v(\r)+v_{hxc}(\r) -\mu] \hat{n}(\r) d\r$ and, as in current-DFT \cite{vignale1987}, we define
\begin{align}
v_{hxc}(\r) &=  \left.\f{\delta \Omega_{hxc}}{\delta n(\r)}\right|_{\phi_{d\k},f_{d\k}} \nn \\
\f{e}{c}\mathbf{A}_{xc}(\r) &= \left.\f{\delta \Omega_{hxc}}{\delta \mathbf{j}_p(\r)}\right|_{\phi_{d\k},f_{d\k}} {.}
\end{align}
Subtraction leads to 
\begin{align}
&(f_{d\k} - f_{d'\k'}) \big< \phi_{d'\k'} \big| \hat{T} + \hat{V} \big| \phi_{d\k} \big> \nn \\ 
&+ \left.\bigg< \phi_{d'\k'} \bigg| \f{\delta \Omega_{hxc}}{\phi_{d\k}^*} \bigg>\right|_{n,\mathbf{j}_p} - \left.\bigg< \f{\delta \Omega_{hxc}}{\phi_{d'\k'}} \bigg| \phi_{d\k} \bigg>\right|_{n,\mathbf{j}_p} = 0 {.} \label{eq:stat:cond:2a}
\end{align} 
The variations of $G$ with respect to $n(\r)$ and $\mathbf{j}_p(\r)$ for fixed $(\phi_{d\k},f_{d\k})$ give
\begin{align}
v(\r) + v_{hxc}(\r) + \f{e^2}{2mc^2} |\mathbf{A}(\r)|^2 - \mu + \left.\f{\delta K_s}{\delta n(\r)}\right|_{\phi_{d\k},f_{d\k}} &= 0  \nn \\
\f{e}{c} \mathbf{A}(\r) + \f{e}{c} \mathbf{A}_{xc}(\r) + \left.\f{\delta K_s}{\delta \mathbf{j}_p(\r)}\right|_{\phi_{d\k},f_{d\k}} &= 0 {.} \label{eq:stat:cond:2b}
\end{align}
The stationary conditions in Eqs.~(\ref{eq:stat:cond:2a}) and (\ref{eq:stat:cond:2b}) are the same as those of the noninteracting system, Eqs.~(\ref{eq:stat:cond:1a}) and (\ref{eq:stat:cond:1b}), if we define 
\begin{align}
v_s(\r) &= v(\r) + v_{hxc}(\r) + \f{e^2}{2mc^2} \big( |\mathbf{A}(\r)|^2 -  |\mathbf{A}_s(\r)|^2 \big)  \nn \\
\mathbf{A}_s(\r) &= \mathbf{A}(\r) + \mathbf{A}_{xc}(\r) 
\label{eq:v0} 
\end{align}
and if we define $\hat{W}_s$ by 
\begin{align}
\langle \phi_{d'\k'} | \hat{W}_s | \phi_{d\k} \rangle = \f{\left.\Big< \phi_{d'\k'} \Big| \f{\delta \Omega_{hxc}}{\delta\phi_{d\k}^*} \Big>\right|_{n,\mathbf{j}_p} - \left.\Big< \f{\delta \Omega_{hxc}}{\delta\phi_{d'\k'}} \Big| \phi_{d\k} \Big>\right|_{n,\mathbf{j}_p}}{f_{d\k}-f_{d'\k'}} \label{eq:w0}
\end{align}
for $d\k\neq d'\k'$.  
A similar expression for a nonlocal effective potential has been derived \cite{pernal2005} in the context of reduced density matrix functional theory at $\tau=0$ \cite{gilbert1975}, where the energy is a functional of all natural orbitals and the orbital derivatives are not constrained to fixed $(n,\mathbf{j}_p)$.  We can obtain the equilibrium $n(\r)$, $\mathbf{j}_p(\r)$, and $\phi_{d\k}(\r)$ by self-consistently solving a single-particle Schr\"odinger equation with the Hamiltonian
\begin{align}
\hat{h}^{\mathit{eff}} = \f{1}{2m}\big(\hat{\mathbf{p}}+\f{e}{c}\mathbf{A}_s(\hat{\r})\big)^2 + \hat{V}_s + \hat{W}_s {,}
\label{eq:GKS:Hamiltonian}
\end{align}
where $\hat{V}_s = \int v_s(\r) \hat{n}(\r) d\r$, together with the expressions for $n(\r)$ and $\mathbf{j}_p(\r)$ in Eq.~(\ref{eq:n}).  The eigenvalues will obey
\begin{align}
\bar{ll}
\epsilon_{n\k} \leq \mu & \quad\textrm{if} \;\; f_{n\k}=1  \\
\epsilon_{d\k} = \mu & \quad\textrm{if} \;\; 0\leq f_{d\k}\leq 1  \\
\epsilon_{n\k} \geq \mu & \quad\textrm{if} \;\; f_{n\k}=0 \ear {.}
\end{align}
\vspace{0.4cm}

\section{Operator orthonormalization}

In Eq.~(\ref{eq:rho:expansion}), the density matrix was expanded in terms of a complete basis of operators $\{\hat{\mu}_i,\hat{\nu}_i,\hat{\xi}_i,\hat{o}_i\}$.  Here, $\{\hat{\mu}_i\}$ is a complete basis of Hermitian one-body operators in the $\mathcal{D}$ subspace that span all operators of the form $c_{d\k}^{\dag} c_{d\k}$, $c_{d\k}^{\dag} c_{d'\k'}+c_{d'\k'}^{\dag} c_{d\k}$, and $-i c_{d\k}^{\dag} c_{d'\k'}+ ic_{d'\k'}^{\dag} c_{d\k}$.  $\{\hat{\nu}_i\}$ is a basis that together with $\{\hat{\mu}_i\}$ provides a complete basis for all Hermitian two-body operators in the $\mathcal{D}$ subspace.   $\{\hat{\mu}_i\}$ and $\{\hat{\nu}_i\}$ are constructed to be orthogonal.

$\{\hat{\xi}_i\}$ is a complementary basis of Hermitian operators that together with $\{\hat{\mu}_i\}$ and $\{\hat{\nu}_i\}$ spans all remaining one- and two-body operators of the form $c_{b_1\k_1}^{\dag} c_{b_1'\k_1'}$ and $c_{b_1\k_1}^{\dag} c_{b_2\k_2}^{\dag} c_{b_2'\k_2'} c_{b_1'\k_1'}$, where $b_i\in\mathcal{S}\cup\mathcal{D}$ and at least one of the indices does not belong to $\mathcal{D}$.  $\{\hat{\xi}_i\}$ is constructed to be orthogonal to $\{\hat{\mu}_i\}$ and $\{\hat{\nu}_i\}$.  $\{\hat{o}_i\}$ is the basis of all remaining operators needed to expand $\hat{\rho}$ and is orthogonal to $\{\hat{\mu}_i\}$, $\{\hat{\nu}_i\}$ and $\{\hat{\xi}_i\}$. 

To construct $\{\hat{\mu}_i\}$, we first define the set $\mathcal{A}$ of all Hermitian one-body operators $\hat{A}_i$ built from operators in $\mathcal{D}$.  The matrix representation of $\hat{A}_i$ in a complete basis of $N$-body determinant states $|D_N\rangle=c_{d_1\k_1}^{\dag} \ldots c_{d_N\k_N}^{\dag}|0\rangle$ is
\begin{align}
M_{Ai} = (\hat{A}_i)_{D_N,D'_N} = \langle D_{N} | \hat{A}_i | D'_{N}\rangle {,}
\end{align}
where $D_N=(d_1\k_1,\ldots,d_N\k_N)$ and $N=1,\ldots,N_{max}$ with $N_{max}$ determined by the 
truncation of the single-particle Hilbert space.  Let $\mathcal{M}_a$ denote the set of matrices $M_{Ai}$ for all $\hat{A}_i\in\mathcal{A}$.  We perform a Gram-Schmidt orthogonalization of $\mathcal{M}_a$ with respect to the Hilbert-Schmidt inner product $\mathrm{Tr}(M_{Ai}^{\dag} M_{Aj})$ to obtain an orthogonalized set $\mathcal{M}^{ortho}_A$.  The linearly independent set $\{\hat{\mu}_i\}$ is then defined by projecting the orthogonalized $\mathcal{M}^{ortho}_A$ back onto $\hat{A}_i$.  If there are linear dependencies among the $\hat{A}_i$, then there will be fewer $\hat{\mu}_i$ than $\hat{A}_i$. 

To proceed, we define the set $\mathcal{B}$ of all Hermitian two-body operators $\hat{B}_i$ built from operators in $\mathcal{D}$ that are not in $\mathcal{A}$; we also define the set $\mathcal{M}_B$ of matrix representations $M_{Bi} = (\hat{B}_i)_{D_N,D'_N}$.  We form the union $\mathcal{M}_{AB}=\mathcal{M}^{ortho}_A \cup \mathcal{M}_B$ and again perform a Gram-Schmidt orthogonalization to obtain the orthogonalized set $\mathcal{M}^{ortho}_{AB}$.  Mapping back onto the operators $\{\hat{\mu}_i,\hat{B}_i\}$ defines a set of operators $\{\hat{\mu}_i,\hat{\nu}_i\}$ with $\hat{\nu}_i$ linearly independent of all $\hat{\mu}_i$.  A similar recursive procedure is used to define the bases $\{\hat{\xi}_i\}$ and $\{\hat{o}_i\}$.  The above procedure has been used to construct operator bases for two-electron states with $S_z=0$ built from a six-dimensional single-particle Hilbert space \cite{requist2014b} and four-electron states with $S_z=0$ built from an eight-dimensional single-particle Hilbert space. 

\bibliography{bibliography2018}

\end{document}